\documentclass{article}
\usepackage{amsfonts}

\usepackage{graphicx,color}
\usepackage{amsmath}

\newenvironment{proof}[1][Proof]{\textbf{#1.} }{\ \rule{0.5em}{0.5em}}
\input{tcilatex}

\definecolor{green}{rgb}{0.00,0.50,0.00}

\begin{document}

\title{Asymptotic Stochastic Transformations for \\
Nonlinear Quantum Dynamical Systems}
\author{John Gough}
\date{}
\maketitle

\begin{abstract}
The Ito and Stratonovich approaches are carried over to quantum stochastic
systems. Here the white noise representation is shown to be the most
appropriate as here the two approaches appear as Wick and Weyl orderings,
respectively. This introduces for the first time the Stratonovich form for
SDEs driven by Poisson processes or quantum SDEs including the conservation
process. The relation of the nonlinear Heisenberg ODES to asymptotic quantum
SDEs is established extending previous work on linear (Schrodinger)
equations. This is shown to generalize the classical integral
transformations between the various forms of stochastic calculi and to
extend the Khasminskii theorem to the quantum setting.
\end{abstract}

\section{Introduction}

The stochastic asymptotic analysis of dynamical systems has its origins in
Einstein's theory of Brownian motion, however, it extends to very general
classes of systems, both classical and quantum.

In the classical case, one considers typically a dynamical variable $x_t
=x_t (\lambda )$ determined from an ODE $\frac{d}{dt} x_t (\lambda ) =
\lambda F(x_t (\lambda ), M_t ), x_0(\lambda ) = x_0$, where $\lambda > 0$
is a coupling parameter, $M_t$ is an external (stochastic) input and $F$ is
some function (usually Lipschitz with bounded first partial derivatives).
Supposing that $\frac{1}{T} \int_t^{T+t}ds \, F(x, M_s)$ converges as $T \to
\infty$ in probability to $F_o(x)$ uniformly in $t$ for arbitrary $x$, then
under suitable conditions [l] the averaging principle states that $%
x_{t/\lambda } (\lambda)$ converges in probability to $\bar{x}_t$, the
solution of the averaged ODE $\frac{d}{dt} \bar{x}_t = F_o( \bar{x}_t); \, 
\bar{x}_0 = x_0.$ One may think of $x_{t/\lambda } (\lambda )$ as a
perturbation away from the averaged solution. If $F_o = 0$, the averaging
principle states that $x_{t/\lambda } (\lambda )$ will not vary
significantly from $x_0$ on time scales $[0, T/\lambda ]$ [2]. This lead
Stratonovich in 1961 to suggest that on time scales $[0, T/\lambda^2 ]$
stochastic fluctuations accumulate appreciably; this idea was substantiated
by Khasminskii [3] who proved that $x_{t/\lambda^2 } (\lambda )$ converges
weakly to a Markov diffusion. In this context, the interpretation in terms
of a white noise Langevin equation, originally given by Wong and Zakai [4],
has been revealing. If $\xi(t)$ are regular stochastic processes with mean
zero and correlation $\mathbb{E} [\xi(t) \xi(s)] = \frac{1}{\lambda^2} c (%
\frac{t-s}{\lambda^2})$, so that the $\lambda \to 0$ limit leads to a white
noise process, then the dynamical equations of motion can converge to an
asymptotic Langevin equation which retains the pre-limit form provided one
uses the Stratonovich version of stochastic calculus [5].

When dealing with quantum systems the scaling employed in the Khasminskii
theorem corresponds to the van Hove re-scaling of the dynamical variables
[6]. The formulations of most quantum Langevin equations encountered in
physics [7] can occasionally be justified as an asymptotic quantum
stochastic limit of such re-scaled variables [S]. The germ of the approach
to be developed in this paper comes from the work of Accardi, Frigerio and
Lu [9] wherein a class of noncommutative Khasminskii theorems are
established. The point of view presented here is closer in spirit to that of
Wong and Zakai. In particular, quantum white noises [10] play a fundamental
role. These are Bose operators (t) satisfying a canonical commutation
relation (CCR) 
\begin{equation}
\lbrack a(t),a(s))]=\kappa \delta _{+}(t-s)+\kappa ^{\ast }\delta _{-}(t-s),
\tag{1.1}
\end{equation}
where $\kappa $ is a complex number and $\delta _{\pm }(t)$ are special
delta functions on the space of regulated functions picking out the
future/past values at a time $t$. The form (1.1) arises as the limit form of
the CCR for physical fields wherein the right-hand side would be typically
the causal (Feynman) propagator. The $a^{\sharp }(t)$ have white noise
spectra and play a role similar to the input processes studied by Gardiner
[11], though with different operational properties and without the effective
restriction $\kappa \equiv \frac{1}{2}$. As such, the choice between Ito and
Stratonovich versions of stochastic calculus corresponds to the choice
between Wick and Weyl ordering, respectively, of the noises with respect to
the integrands. Note that the product of two Wick ordered expressions will
require further re-ordering to attain Wick order again, this leads to extra
terms which correspond to the standard shift accounted for by the quantum
Ito table [12]. This notion, in fact, dates back to an observation of Hudson
and eater [ 131. It should also be mentioned that an alternate description
of Stratonovich quantum stochastic calculus, not involving white noise and
taking $\kappa =\frac{1}{2}$, has been given independently by Chebotarev
[14]. It is worth mentioning that while the stochastic limits described here
should preserve canonical structures, especially unitarity, many important
physical properties such as detailed balance and the KMS condition are
typically lost [15]. Heuristic accounts of these white noise processes have
appeared [16, 171 along with an account of their applications to physic
[18]. However, only linear systems have been treated up to now. In this
paper, the analysis is extended to nonlinear systems which essentially
emerge as Heisenberg evolutes corresponding to linear quantum SDEs (i.e.
stochastic Schrodinger equations). Complete transformation laws are derived.
The classical formulae for Ito-Stratonovich conversion for both Wiener and
Poisson driven systems are recovered as special cases.

\section{Classical stochastic differential equations}

In this section several well-known features [19] of SDEs are reviewed. Let $%
u,\sigma ^{(\alpha )}\in C(^{n}\times \mathbb{R},\mathbb{R}^{n})$ satisfy
appropriate Lipschitz and growth conditions, and let $M_{t}^{(\alpha )}$ be
a set of stochastic processes on a probability space $\left( \Omega ,%
\mathcal{F},\mathbb{P}\right) $. Let $x_{0}\in L^{2}\left( \Omega ,\mathcal{F%
},\mathbb{P}\right) $ be independent of the processes $M_{t}^{(\alpha )}$
for $t>0$. Whenever the $M_{t}^{(\alpha )}$ are nondifferentiable, the
formal equation (2.1) below is technically meaningless, 
\begin{equation}
\frac{dX_{t}}{dt}=v\left( X_{t},t\right) +\sum_{\alpha }\sigma ^{\left(
\alpha \right) }\left( X_{t},t\right) \frac{dM_{t}^{\left( \alpha \right) }}{%
dt},\quad X_{0}=x_{0}.  \tag{2.1}
\end{equation}
Different finite-step numerical schemes used to define an estimate for $X$,
in (2.1) lead to inconsistencies in the small time-step limit which do not
arise when dealing with ODES. For instance, let $\mathcal{P}$ be an ordered
partition of $[0,t]$, say $[0=t_{0}<t_{1}<\cdots <t_{N}=t]$, and denote $%
\text{max}|t_{j+1}-t_{j}|$ by $|\mathcal{P}|$. One constructs an approximate
solution $X_{T}^{\mathcal{P}}$ to (2.1) in the form of a random variable $%
X_{t_{N}}^{\mathcal{P}}$, obtained by the iterative scheme 
\begin{equation}
X_{t_{j+1}}^{\mathcal{P}}=X_{t_{j}}^{\mathcal{P}}+\Delta X_{t_{j}}^{\mathcal{%
P}},\quad X_{t_{0}}^{\mathcal{P}}=0.  \tag{2.2}
\end{equation}
One then understands the limit of $X_{T}^{\mathcal{P}}$ for finer partitions
as a mean-square limit, 
\begin{equation}
X_{t}\equiv x_{0}+\int_{0}^{t}\left\{ v\left( X_{s},s\right) ds+\sum_{\alpha
}\sigma ^{\left( \alpha \right) }\left( X_{s},s\right) dM_{s}^{\left( \alpha
\right) }\right\} :=\underset{|\mathcal{P}!\rightarrow 0}{\text{l.i.m.}}%
X_{t_{N}}^{\mathcal{P}}.  \tag{2.3}
\end{equation}
That is, $\lim_{|\mathcal{P}!\rightarrow 0}\mathbb{E}\left[ \left|
X_{t}-X_{t_{N}}^{\mathcal{P}}\right| ^{2}\right] =0$ if such a limit $X_{t}^{%
\mathcal{P}}\in L^{2}\left( \Omega ,\mathcal{F},\mathbb{P}\right) $ exists.
Amongst the various schemes to compute the increment $\Delta X_{t_{j}}^{%
\mathcal{P}}$, the simplest is the Euler method 
\begin{equation}
\Delta X_{t_{j}}^{\mathcal{P}}:=v\left( X_{t_{j}}^{\mathcal{P}},t_{j}\right) %
\left[ t_{j+1}-t_{j}\right] +\sum_{\alpha }\sigma ^{\left( \alpha \right)
}\left( X_{t_{j}}^{\mathcal{P}},t_{j}\right) \left[ M_{t_{j+1}}^{\left(
\alpha \right) }-M_{t_{j}}^{\left( \alpha \right) }\right] ,  \tag{2.4}
\end{equation}
an alternative is given by the averaging scheme 
\begin{eqnarray}
\Delta X_{t_{j}}^{\mathcal{P}} &:&=\frac{1}{2}\left[ v\left( X_{t_{j+1}}^{%
\mathcal{P}},t_{j+1}\right) +v\left( X_{t_{j}}^{\mathcal{P}},t_{j}\right) %
\right] \left[ t_{j+1}-t_{j}\right]  \notag \\
&&+\sum_{\alpha }\left[ \sigma ^{\left( \alpha \right) }\left( X_{t_{j+1}}^{%
\mathcal{P}},t_{j+1}\right) +\sigma ^{\left( \alpha \right) }\left(
X_{t_{j}}^{\mathcal{P}},t_{j}\right) \right] \left[ M_{t_{j+1}}^{\left(
\alpha \right) }-M_{t_{j}}^{\left( \alpha \right) }\right] .  \TCItag{2.5}
\end{eqnarray}
If the $\sigma ^{(\alpha )}$ were zero, so that (2.1) reduces to an ODE,
then the Euler scheme has local error of $O(h^{2})$, whereas the averaging
scheme has local error of $O(h^{3})$, where $h$ is the largest increment
size (which is, of course, proportional to $|\mathcal{P}|$ for the
deterministic case). If the $\sigma ^{(\alpha )}$ are nonzero then the noise
increments $M_{t_{j+1}}-M_{t_{j}}$ may fluctuate rapidly enough to lead to $%
h=O(|\mathcal{P}|^{a})$ with $a<1$. For instance, in the case of the Wiener
process as noise one has $a=\frac{1}{2}$. In this situation the Euler scheme
leads to an $0(|\mathcal{P}|)$ local error which makes a nontrivial
contribution to the stochastic integral (2.3) which is not present in the
averaging scheme. The respective solutions (when they exist) are called the
Ito and Stratonovich stochastic integrals and the following notation shall
be used: $X_{t}=X_{0}+\int_{0}^{t}dX_{s}$, with 
\begin{eqnarray}
\text{(Ito)\quad }dX_{t} &=&v\left( X_{t},t\right) dt+\sum_{\alpha }\sigma
^{\left( \alpha \right) }\left( X_{t},t\right) dM_{t}^{\left( \alpha \right)
},  \TCItag{2.6a} \\
\text{(Stratonovich)\quad }dX_{t} &=&\overline{v\left( X_{t},t\right) }%
dt+\sum_{\alpha }\overline{\sigma ^{\left( \alpha \right) }\left(
X_{t},t\right) }dM_{t}^{\left( \alpha \right) }.  \TCItag{2.6b}
\end{eqnarray}
The deterministic component of the integrand (coefficient of $dt$) is taken
to be Riemann integrable and so the over- or under-bar notation can be
omitted, as frequently will be the case. In general, for $X_{t}$ and $Y_{t}$
stochastic processes, one understands the limits $X_{t_{j}}^{{}},\quad =0.$ 
\begin{equation}
\int_{0}^{t}X_{s}dY_{s}:=\underset{\left| \mathcal{P}\right| \rightarrow 0}{%
\text{l.i.m.}}\sum_{j=0}^{N-1}X_{t_{j}}^{\mathcal{P}}\Delta Y_{t_{j}}^{%
\mathcal{P}},\quad \int_{0}^{t}\overline{X_{s}}dY_{s}:=\underset{\left| 
\mathcal{P}\right| \rightarrow 0}{\text{l.i.m.}}\sum_{j=0}^{N-1}\overline{%
X_{t_{j}}^{\mathcal{P}}}\Delta Y_{t_{j}}^{\mathcal{P}},  \tag{2.7}
\end{equation}
where $\Delta Y_{t_{j}}^{\mathcal{P}}:=Y_{t_{j+1}}^{\mathcal{P}}-Y_{t_{j}}^{%
\mathcal{P}}$ and $\overline{X_{t_{j}}^{\mathcal{P}}}:=\frac{1}{2}\left(
X_{t_{j+1}}^{\mathcal{P}}-X_{t_{j}}^{\mathcal{P}}\right) $. Moreover, to
compute the increments of the product $X,Y$, one notes 
\begin{eqnarray}
\Delta \left( XY\right) _{t_{j}}^{\mathcal{P}} &=&X_{t_{j+1}}^{\mathcal{P}%
}Y_{t_{j+1}}^{\mathcal{P}}-X_{t_{j}}^{\mathcal{P}}Y_{t_{j}}^{\mathcal{P}%
}=X_{t_{j}}^{\mathcal{P}}\Delta Y_{t_{j}}^{\mathcal{P}}+\Delta X_{t_{j}}^{%
\mathcal{P}}Y_{t_{j}}^{\mathcal{P}}+\Delta X_{t_{j}}^{\mathcal{P}}\Delta
Y_{t_{j}}^{\mathcal{P}}  \notag \\
&=&\overline{X_{t_{j}}^{\mathcal{P}}}\Delta Y_{t_{j}}^{\mathcal{P}}+\Delta
X_{t_{j}}^{\mathcal{P}}\overline{Y_{t_{j}}^{\mathcal{P}}}.  \TCItag{2.8}
\end{eqnarray}
The terms $\Delta X\Delta Y$ need not be negligible; for the Wiener process
it is $O(|\mathcal{P}|)$. Thus, if the noise processes $M_{t}^{(\alpha )}$
admit a stochastic calculus and if $X_{t},Y_{t}$ are processes driven by
these noises, then one expects that the Ito form breaks the Leibniz rule of
calculus, 
\begin{equation}
d\left( X_{t}Y_{t}\right) =X_{t}dY_{t}+dX_{t}Y_{t}+dX_{t}dY_{t},  \tag{2.9a}
\end{equation}
while the Stratonovich form retains it 
\begin{equation}
d\left( X_{t}Y_{t}\right) =\overline{X_{t}}dY_{t}+dX_{t}\overline{Y_{t}}; 
\tag{2.9b}
\end{equation}
here the equivalence relation means equality up to $O(dt)$ in the
deterministic part which implies equality under the integral sign. Using
(2.9b) inductively, one sees that 
\begin{equation}
dX_{t}^{2}=2\overline{X_{t}}dX_{t},\quad dX_{t}^{3}=\left( 2\overline{X_{t}}%
^{2}+\overline{X_{t}^{2}}\right) dX_{t},\quad dX_{t}^{4}=\left( \overline{%
X_{t}^{3}}+\overline{X_{t}}\overline{X_{t}^{2}}+2\overline{X_{t}}^{3}\right)
dX_{t},\quad \cdots .  \tag{2.10}
\end{equation}
Now $\overline{X_{t}^{n}}=\frac{1}{2}\left( X_{t}+dX_{t}\right) ^{n}+\frac{1%
}{2}X_{t}^{n}$, from which it follows that $\overline{X_{t}^{n}}-\overline{%
X_{t}}^{n}=O\left( \left( dX_{t}\right) ^{2}\right) $ and so 
\begin{equation*}
\overline{X_{t}^{n}}dX_{t}=\overline{X_{t}}^{n}dX_{t}+o\left( \left(
dX_{t}\right) ^{2}\right) .
\end{equation*}
In the case of diffusion processes, one then has $\overline{X_{t}^{n}}%
dX_{t}\equiv \overline{X_{t}}^{n}dX_{t}$, and so $dX_{t}^{n}\equiv n%
\overline{X_{t}}^{n-1}dX_{t}$ which implies that the chain rule formula of
standard calculus holds when applying the Stratonovich calculus to
ditfusions. However, one can see already that this will not be true for
processes driven by the Poisson process with nonlinear noise coefficients.

\bigskip

\noindent \textbf{Examples} Whereas the Stratonovich integral with respect
to general martingales can be formulated, it is interesting to give
explicitly the Wiener and Poisson cases.

\bigskip

\noindent \textbf{Wiener Process as noise:} The Wiener process $B_{t}$,
admits a stochastic calculus with $(dB_{t})^{2}=dt$ and higher powers $o(dt)$%
. The Stratonovich SDE $dX_{t}=v(X_{t},t)dt+\overline{\sigma (X_{t},t)}%
dB_{t} $ is well known to be equivalent to an Ito SDE $dX_{t}=\tilde{v}%
(X_{t},t)dt+\tilde{\sigma}(X_{t},t)dB_{t}$, where the coefficients are
related by 
\begin{equation}
\tilde{v}=v+\frac{1}{2}\sigma \sigma ^{\prime },\quad \tilde{\sigma}=\sigma ,
\tag{2.11}
\end{equation}
with $\sigma ^{\prime }(x,t):=\frac{\partial }{\partial x}\sigma (x,t)$.
Inversely, $v=\tilde{v}-\frac{1}{2}\tilde{\sigma}^{\prime }\tilde{\sigma}$.

\bigskip

\noindent \textbf{Poisson process as noise:} The Poisson process Nt admits a
stochastic calculus with $(dN_{t})^{n}=dN$, for all integers $n>_{1}$. The
fluctuations in $dN_{t}$ are $0(dt)$. The Stratonovich SDE $dX_{t}=\overline{%
v(X_{t},t)}dt+\overline{\mu (X_{t},t)}dN_{t}$, is equivalent to the Ito SDE $%
dX_{t}=\tilde{v}(X_{t},t)dt+\tilde{\mu}(X_{t},t)dN_{t}$, where the
coefficients are related by 
\begin{equation}
\tilde{v}=v,\quad \tilde{\mu}\left( x,t\right) =\frac{1}{2}\left[ \mu \left(
x+\tilde{\mu}\left( x,t\right) \right) +\mu \left( x,t\right) \right] . 
\tag{2.12}
\end{equation}
The Stratonovich form for the Poisson integrals is next derived explicitly.
Let $f=f(x)$ be analytic, then by Taylor expansion one has 
\begin{eqnarray*}
f\left( X_{t_{j+1}}^{\mathcal{P}}\right) &=&f\left( X_{t_{j}}^{\mathcal{P}%
}\right) +f^{\prime }\left( X_{t_{j}}^{\mathcal{P}}\right) \Delta X_{t_{j}}^{%
\mathcal{P}}+\frac{1}{2!}f^{\prime \prime }\left( X_{t_{j}}^{\mathcal{P}%
}\right) \left( \Delta X_{t_{j}}^{\mathcal{P}}\right) ^{2}+\cdots \\
&\equiv &f\left( X_{t_{j}}^{\mathcal{P}}\right) +\left\{ f^{\prime }\left(
X_{t_{j}}^{\mathcal{P}}\right) \tilde{\mu}\left( X_{t_{j}}^{\mathcal{P}%
},t_{j}\right) +\frac{1}{2!}f^{\prime \prime }\left( X_{t_{j}}^{\mathcal{P}%
}\right) \tilde{\mu}\left( X_{t_{j}}^{\mathcal{P}},t\right) +\cdots \right\}
\Delta N_{t_{j}} \\
&&+f^{\prime }\left( X_{t_{j}}^{\mathcal{P}}\right) \tilde{v}\left(
X_{t_{j}}^{\mathcal{P}},t_{j}\right) \Delta t_{j}+o\left( \Delta
t_{j}\right) .
\end{eqnarray*}
Therefore, multiplying by $\Delta N_{t_{j}}=N_{t_{j+1}}-N_{t_{j}}$, and
re-summing leads to 
\begin{equation*}
f\left( X_{t_{j}}^{\mathcal{P}}\right) \Delta N_{t_{j}}\equiv f\left(
X_{t_{j}}^{\mathcal{P}}+\tilde{\mu}\left( X_{t_{j}}^{\mathcal{P}%
},t_{j}\right) \right) \Delta N_{t_{j}}
\end{equation*}
On account of this, 
\begin{eqnarray*}
X_{t_{j+1}}^{\mathcal{P}} &=&X_{t_{j}}^{\mathcal{P}}+\frac{1}{2}\left\{
v\left( X_{t_{j+1}}^{\mathcal{P}},t_{j+1}\right) +v\left( X_{t_{j}}^{%
\mathcal{P}},t_{j}\right) \right\} \Delta t_{j} \\
&&+\frac{1}{2}\left\{ \mu \left( X_{t_{j+1}}^{\mathcal{P}},t_{j+1}\right)
+\mu \left( X_{t_{j}}^{\mathcal{P}},t_{j}\right) \right\} \Delta N_{t_{j}} \\
&\equiv &X_{t_{j}}^{\mathcal{P}}+v\left( X_{t_{j}}^{\mathcal{P}%
},t_{j}\right) \Delta t_{j}+\frac{1}{2}\left\{ \mu \left( X_{t_{j}}^{%
\mathcal{P}}+\tilde{\mu}\left( X_{t_{j}}^{\mathcal{P}},t_{j}\right)
,t_{j}\right) +\mu \left( X_{t_{j}}^{\mathcal{P}},t_{j}\right) \right\}
\Delta N_{t_{j}}.
\end{eqnarray*}
and so the relation (2.12) closes the transformation.

\bigskip

\noindent \textbf{Remarks} i) In the case of linear Poisson-noise
coefficient $\mu (x)=\mu _{0}x+c$ the relation (2.12) gives $\tilde{\mu}%
(x)=(\mu _{0}x+c)/(l-\frac{1}{2}\mu _{0})$. The requirement on the
coefficients is that $x\mapsto x+\mu (x,t)$ is strictly monotone, for each $%
t $, and so defines a proper change of variable. For general $\mu (x)$ there
exists no closed expression for $\tilde{\mu}$, however, it may be
approximated by the iteration $\tilde{\mu}_{n+1}(x)=\frac{1}{2}\{\mu (x+%
\tilde{\mu}_{n}(x))+\mu (x)\},\mu _{0}(x)=\mu (x)$.

ii) There are other numerical schemes which can be employed. For instance,
the mid-point 2nd order Runge-Kutta method 
\begin{equation*}
X_{t_{j+1}}^{\mathcal{P}}=X_{t_{j}}^{\mathcal{P}}+v\left( \tilde{X}%
_{t_{j}^{\ast }}^{\mathcal{P}},t_{j}^{\ast }\right) \Delta
t_{j}+\sum_{\alpha }\sigma ^{\left( \alpha \right) }\left( \tilde{X}%
_{t_{j}^{\ast }}^{\mathcal{P}},t_{j}^{\ast }\right) \Delta M_{t_{j}}^{\left(
\alpha \right) },
\end{equation*}
where $t_{j}^{\ast }:=\frac{1}{2}\left( t_{j+1}+t_{j}\right) $ and $\tilde{X}%
_{t_{j}^{\ast }}^{\mathcal{P}}$ is obtained from an Euler time-step from $%
t_{j}$ to $t_{j}^{\ast }$ 
\begin{equation*}
\tilde{X}_{t_{j}^{\ast }}^{\mathcal{P}}=X_{t_{j}}^{\mathcal{P}}+v\left( 
\tilde{X}_{t_{j}^{\ast }}^{\mathcal{P}},t_{j}^{\ast }\right) [t_{j}^{\ast
}-t_{j}]+\sum_{\alpha }\tilde{\sigma}^{\left( \alpha \right) }\left(
X_{t_{j}}^{\mathcal{P}},t_{j}^{\ast }\right) [M_{t_{j}^{\ast }}^{\left(
\alpha \right) }-M_{t_{j}}^{\left( \alpha \right) }]
\end{equation*}
with $\tilde{v},\tilde{\sigma}^{(\alpha )}$ the associated Ito coefficients.
This procedure is, in fact, equivalent to the Stratonovich choice. However,
the method given by is equivalent to the Stratonovich choice only for the
Wiener process as noise. For the Poisson process, it coincides with the
Stratonovich choice only if $\mu (x,t)$ is linear in $x$; otherwise it leads
to a stochastic integral equivalent to an Ito integral with coefficient $%
\hat{\mu}(x,t)$ related by $\mu (x+\frac{1}{2}\hat{\mu}(x,t),t)=\hat{\mu}%
(x,t)$.

iii) In general $X_{t}$ is said to be Riemann integrable with respect to $%
Y_{t},$ if the mean-square limit of $\sum_{j}X_{s_{j}}^{\mathcal{P}}\Delta
Y_{t_{j}}^{\mathcal{P}}$ exists whenever an arbitrary prescription to
determine $s\in \lbrack t_{j},t_{j+l}]$ is given. The Ito calculus is based
on the choice $s_{j}=t_{j}$. One may call this the \emph{retarded} Ito
theory. The \emph{advanced} Ito theory can be defined via the prescription
that $s_{j}=t_{j+l}$. The Stratonovich calculus then corresponds to the
averaging over the retarded and advanced Ito versions. An alternative
definition of Stratonovich integrals as simply mean-square Riemann integrals
with $s_{j}=t_{j}^{\ast }$ is frequently taken in probabilistic literature,
however, as seen in the previous remark this does not always coincide with
the definition here.

\section{Quantum Stochastic Calculus}

We begin by recalling the Hudson-Parthasarathy theory of noncommutative
stochastic processes [12, 20]. Let $\mathcal{H}_{0}$ and $\mathcal{K}$ be
fixed Hilbert spaces called the initial and internal spaces, respectively. $%
L^{2}(I;\mathcal{K})\cong L^{2}$ $\left( I\right) \otimes \mathcal{K}$ is
the Hilbert space of square-integrable $\mathcal{K}$-valued functions on an
interval $I\subset \mathbb{R}^{+}$. The (Bose) Fock space over $L^{2}(I;%
\mathcal{K})$ is denoted by $\Gamma _{+}(I;\mathcal{K})$. In the following,
a quantum stochastic process will be understood as an operator-valued family 
$(X_{t})_{t\geq 0}$ on $\mathcal{H}_{0}$ $\otimes \Gamma _{+}(\mathbb{R}^{+},%
\mathcal{K})$. \ The initial value of the process $X_{0}$ will be taken as
an element $x_{0}\in \mathcal{B}(\mathcal{H}_{0})$. The noise space admits
the following continuous tensor-product decomposition, 
\begin{equation}
\Gamma _{+}(\mathbb{R}^{+},\mathcal{K})\cong \Gamma _{+}(\left[ 0,t\right] ;%
\mathcal{K})\otimes \Gamma _{+}(\left( t,\infty \right) ;\mathcal{K}), 
\tag{3.1}
\end{equation}
which allows the introduction of a time filtration. In particular, a process 
$\left( X_{t}\right) _{t\geq 0}$ is adapted if, for each $t>0$, $%
X_{t}\varphi =\varphi $, for all $\varphi \in \Gamma _{+}(\left( t,\infty
\right) ;\mathcal{K})$.

Let $\Psi \left( f\right) $ denote the exponential vector with test function 
$f\in L^{2}(\mathbb{R}^{+},\mathcal{K})$ and for some subset $\mathcal{S}$
let EXP($\mathcal{S}$) be the span of exponential vectors with test
functions in $\mathcal{S}$; the conserver $\Lambda \left( M\right) $,
creator $A^{\dag }\left( g\right) $ and destroyer $A(g)$ are operators on $%
\Gamma _{+}(\mathbb{R}^{+},\mathcal{K})$ defined by their actions on the
span of exponential vectors, 
\begin{eqnarray}
\Lambda \left( M\right) \Psi \left( f\right) &:&=-i\left. \frac{\partial }{%
\partial \varepsilon }\Psi \left( e^{i\varepsilon M}f\right) \right|
_{\varepsilon =0},  \notag \\
A\left( g\right) \Psi \left( f\right) &:&=\left. \frac{\partial }{\partial
\varepsilon }\Psi \left( f+\varepsilon g\right) \right| _{\varepsilon =0}, 
\notag \\
A^{\dag }\left( g\right) \Psi \left( f\right) &:&=\langle g,f\rangle \Psi
\left( g\right) .  \TCItag{3.2}
\end{eqnarray}
The following notation will be used here 
\begin{equation}
A_{t}^{11}:=\Lambda \left( P_{\left[ 0,t\right] }\right) ,\;A_{t}^{10}\left(
g\right) :=A^{\dag }\left( \chi _{\left[ 0,t\right] }\underline{\otimes }%
g\right) ,\;A_{t}^{01}\left( g\right) :=A\left( \chi _{\left[ 0,t\right] }%
\underline{\otimes }g\right) ,\;A_{t}^{00}:=t,  \tag{3.3}
\end{equation}
for $t\geq 0$, $g\in L_{\text{loc}}^{\infty }\left( \mathbb{R}^{+};\mathcal{K%
}\right) $ and $P_{\left[ 0,t\right] }:f\mapsto \chi _{\left[ 0,1\right] }%
\underline{\otimes }f$. They are called the conserver, creator, destroyer
and standard time process, respectively. One may write $%
A_{t}^{11},A_{t}^{10},A_{t}^{01},A_{t}^{00}$ for the creator and destroyer
when the intensity $g\equiv 1$. For general $g$ one then has $%
A_{t}^{10}\left( g\right) =\int_{0}^{t}g\left( s\right) dA_{s}^{10}$ and $%
A_{t}^{01}\left( g\right) =A_{t}^{10}\left( g\right) ^{\dag }$.

Let $\mathcal{D}$ be a linear domain in $\mathcal{H}_{0}$ and let $\mathcal{S%
}$ be a dense linear manifold (called an admissible space) in $L^{2}\left( 
\mathbb{R}^{+};\mathcal{K}\right) $ closed under the action of the
projections $P_{\left[ 0,t\right] }$, $\left( t\geq 0\right) $. An adapted
process $\left( X_{t}\right) _{t\geq 0}$ is said to be based on $\left( 
\mathcal{D},\mathcal{S}\right) $ if each $X_{t}$, is the ampliation to $%
\mathcal{D}\otimes \mathrm{EXP}\left( P_{\left[ 0,t\right] }\mathcal{S}%
\right) \otimes \Gamma _{+}\left( \left( t,\infty \right) ;\mathcal{K}%
\right) $of an operator on the domain $\mathcal{D}\otimes \mathrm{EXP}\left(
P_{\left[ 0,t\right] }\mathcal{S}\right) $.

Hudson and Parthasarathy [12] define stochastic integrals of the type 
\begin{eqnarray}
X_{t} &:&=x_{0}+\int_{0}^{t}\left\{ X_{s}^{11}\otimes
dA_{s}^{11}+X_{s}^{10}\otimes dA_{s}^{10}+X_{s}^{01}\otimes
dA_{s}^{01}+X_{s}^{00}\otimes dA_{s}^{00}\right\}  \notag \\
&=&x_{0}+\int_{0}^{t}X_{s}^{\alpha \beta }\otimes dA_{s}^{\alpha \beta }, 
\TCItag{3.4}
\end{eqnarray}
where the $\left( X_{t}^{\alpha \beta }\right) _{t\geq 0}$ are adapted
quantum stochastic processes based on $\left( \mathcal{D},\mathcal{S}\right) 
$ which are weakly measurable and satisfy the local square-integrability
conditions below, 
\begin{equation}
\int_{0}^{t}ds\left| f\left( s\right) \right| \left\| X_{s}^{11}u\underline{%
\otimes }\Psi \left( f\right) \right\| ^{2}<\infty ,\;\int_{0}^{t}ds\left\|
X_{s}^{\alpha \beta }u\underline{\otimes }\Psi \left( f\right) \right\|
^{2}<\infty ,  \tag{3.5}
\end{equation}
for arbitrary $t>0$, $u\in \mathcal{D}$ and $f\in \mathcal{S}$.

The notation in (3.4) disguises the fact that the integrators commute with
the integrands (when adapted). The algebraic manipulation of stochastic
integrals is then summarized by the quantum Ito formula for adapted quantum
stochastic processes $X_{t}$ and $Y_{t}$,

\begin{equation}
d(X_{t}Y_{t})=X_{t}\;dY_{t}+dX_{t}\;Yt+dX_{t}\;dY_{t}  \tag{3.6}
\end{equation}
where the last term is the quantum Ito correction and can be evaluated using
the quantum Ito table 
\begin{equation}
dA_{t}^{\alpha 1}dA_{t}^{1\beta }=dA_{t}^{\alpha \beta },\text{ all other
products vanishing.}  \tag{3.7}
\end{equation}

\noindent \textbf{DEFINITION 3.1.} Let $X_{t}$ and $Y_{t}$ be adapted
processes. One defines \emph{Stratonovich quantum stochastic integrals} as

\begin{equation}
\int \overline{X_{t}}dY_{t}:=\int \left( X_{t}+\frac{1}{2}dX_{t}\right)
dY_{t};\int dX_{t}\overline{Y_{t}}:=\int dX_{t}\left( Y_{t}+\frac{1}{2}%
dY_{t}\right) \;  \tag{3.8}
\end{equation}

The product formula is then $X_{t}Y_{t}=x_{0}y_{0}+\int_{0}^{t}\left( dX_{s}%
\overline{Y_{s}}+\overline{X_{s}}dY_{s}\right) $. The integrands $\overline{%
X_{t}}$, $\overline{Y_{t}}$ do not commute with the differentials $%
dX_{t},dY_{t}$ under the integral sign. This lack of commutativity will
result in Stratonovich integrals having left and right hand forms. To work
out general transformations between the two calculi for functions of a
process, the following lemma will be useful. It is easily established by
induction.

\noindent \textbf{LEMMA 3.2.} If $X_{t}$ satisjies the QSDE $%
dX_{t}=X_{t}^{\alpha \beta }\otimes dA_{t}^{\alpha \beta }$ then, for
polynomial $f=f(x)=\sum f_{n}x^{n}$, the quantum Ito calculus gives 
\begin{equation}
df\left( X_{t}\right) =f\left( X_{t}+dX_{t}\right) -f\left( X_{t}\right)
=f\left( X_{t}\right) ^{\alpha \beta }\otimes dA_{t}^{\alpha \beta }, 
\tag{3.9}
\end{equation}
where 
\begin{eqnarray}
f\left( X_{t}\right) ^{11} &=&f\left( X_{t}+X_{t}^{11}\right) -f\left(
X_{t}\right) ,  \notag \\
f\left( X_{t}\right) ^{10} &=&\sum_{n}f_{n}\sum_{p_{1},p_{2}}^{n}\left(
X_{t}+X_{t}^{11}\right) ^{p_{1}}X_{t}^{10}\left( X_{t}\right) ^{p_{2}}, 
\notag \\
f\left( X_{t}\right) ^{01} &=&\sum_{n}f_{n}\sum_{p_{1},p_{2}}^{n}\left(
X_{t}\right) ^{p_{1}}X_{t}^{01}\left( X_{t}+X_{t}^{11}\right) ^{p_{2}}, 
\notag \\
f\left( X_{t}\right) ^{00} &=&\sum_{n}f_{n}\sum_{p_{1},p_{2}}^{n}\left(
X_{t}\right) ^{p_{1}}X_{t}^{00}\left( X_{t}\right) ^{p_{2}}  \notag \\
&&+\sum_{n}f_{n}\sum_{p_{1},p_{2},p_{3}}^{n}\left( X_{t}\right)
^{p_{1}}X_{t}^{01}\left( X_{t}+X_{t}^{11}\right) ^{p_{2}}X_{t}^{10}\left(
X_{t}\right) ^{p_{3}},  \TCItag{3.10}
\end{eqnarray}

Now, from the definition $\overline{f\left( X_{t}\right) }dA_{t}^{\alpha
\beta }=\left[ f\left( X_{t}\right) +\frac{1}{2}df\left( X_{t}\right) \right]
dA_{t}^{\alpha \beta }$ and with the use of the quantum Ito table, one has 
\begin{eqnarray}
\overline{f\left( X_{t}\right) }dA_{t}^{11} &=&\frac{1}{2}\left[ f\left(
X_{t}+X_{t}^{11}\right) +f\left( X_{t}\right) \right] dA_{t}^{11}+\frac{1}{2}%
f\left( X_{t}\right) ^{01}dA_{t}^{01},  \notag \\
\overline{f\left( X_{t}\right) }dA_{t}^{10} &=&\frac{1}{2}\left[ f\left(
X_{t}+X_{t}^{11}\right) +f\left( X_{t}\right) \right] dA_{t}^{10}+\frac{1}{2}%
f\left( X_{t}\right) ^{01}dA_{t}^{00},  \notag \\
\overline{f\left( X_{t}\right) }dA_{t}^{01} &=&f\left( X_{t}\right)
dA_{t}^{01},  \notag \\
\overline{f\left( X_{t}\right) }dA_{t}^{00} &=&f\left( X_{t}\right)
^{00}dA_{t}^{00}.  \TCItag{3.11}
\end{eqnarray}

Likewise

\begin{eqnarray}
dA_{t}^{11}\overline{f\left( X_{t}\right) } &=&\frac{1}{2}\left[ f\left(
X_{t}+X_{t}^{11}\right) +f\left( X_{t}\right) \right] dA_{t}^{11}+\frac{1}{2}%
f\left( X_{t}\right) ^{01}dA_{t}^{01},  \notag \\
dA_{t}^{10}\overline{f\left( X_{t}\right) } &=&f\left( X_{t}\right)
dA_{t}^{10},  \notag \\
dA_{t}^{01}\overline{f\left( X_{t}\right) } &=&\frac{1}{2}\left[ f\left(
X_{t}+X_{t}^{11}\right) +f\left( X_{t}\right) \right] dA_{t}^{01}+\frac{1}{2}%
f\left( X_{t}\right) ^{10}dA_{t}^{00},  \notag \\
dA_{t}^{00}\overline{f\left( X_{t}\right) } &=&f\left( X_{t}\right)
^{00}dA_{t}^{00}.  \TCItag{3.11}
\end{eqnarray}

Remember though that, as in the classical situation, these notations apply
only under the integral sign and are strictly nonassociative; that is, $%
\overline{X_{t}Y_{t}}dZ_{t}\neq \overline{X_{t}}\,\overline{Y_{t}}dZ_{t}$.

\bigskip

\noindent \textbf{THEOREM 3.3.} The following transformations exist between
QSDEs where the coefficients are related according to (2.11) and (2.12),
respectively, 
\begin{eqnarray*}
(i)\quad dX_{t} &=&\overline{v\left( X_{t},t\right) }dt+\overline{\sigma
\left( X_{t},t\right) }\left\{ A_{t}^{10}+dA_{t}^{10}\right\} \\
&=&\tilde{v}\left( X_{t},t\right) dt+\tilde{\sigma}\left( X_{t},t\right)
\left\{ A_{t}^{10}+dA_{t}^{10}\right\} \\
&=&dt\,\overline{v\left( X_{t},t\right) }dt+\left\{
A_{t}^{01}+dA_{t}^{01}\right\} \overline{\sigma \left( X_{t},t\right) }, \\
(ii)\quad dX_{t} &=&\overline{v\left( X_{t},t\right) }dt+\overline{\mu
\left( X_{t},t\right) }\left\{
dA_{t}^{11}+dA_{t}^{10}+dA_{t}^{10}+dA_{t}^{00}\right\} \\
&=&\tilde{v}(X_{t},t)dt+\tilde{\mu}\left( X_{t},t\right) \left\{
dA_{t}^{11}+dA_{t}^{10}+dA_{t}^{10}+dA_{t}^{00}\right\} \\
&=&dt\,\overline{v\left( X_{t},t\right) }+\left\{
dA_{t}^{11}+dA_{t}^{10}+dA_{t}^{10}+dA^{00}\right\} \overline{\mu \left(
X_{t},t\right) }
\end{eqnarray*}

\begin{proof}
In (i), the process $A_{t}^{10}+A_{t}^{01}$ in the Fock vacuum state,
corresponds to the Wiener process. Here $X_{t}^{11}=0$, $%
X_{t}^{01}=X_{t}^{10}=\tilde{\sigma}\left( X_{t},t\right) $ and $X_{t}^{00}=%
\tilde{v}\left( X_{t},t\right) $. From (3.11) one has that the reordering of 
$dA_{t}^{10}$ will lead to the Stratonovich QSDE with integrators to the
right being related to the Ito QSDE by 
\begin{equation}
\overline{f\left( X_{t}\right) }dA_{t}^{10}=f\left( X_{t}\right) dA_{t}^{10}+%
\frac{1}{2}\sum_{n}f_{n}n\left( X_{t}\right) ^{n-1}\tilde{\sigma}\left(
X_{t},t\right) dA_{t}^{00},  \tag{3.13}
\end{equation}
the last term is clearly just $\frac{1}{2}f^{\prime }\left( X_{t}\right) 
\tilde{\sigma}\left( X_{t},t\right) dt$ and so 
\begin{equation}
\overline{\tilde{\sigma}\left( X_{t},t\right) }dA_{t}^{10}=\tilde{\sigma}%
\left( X_{t},t\right) dA_{t}^{10}+\frac{1}{2}\tilde{\sigma}^{\prime }\left(
X_{t},t\right) \tilde{\sigma}\left( X_{t},t\right) dA_{t}^{00}  \tag{3.14}
\end{equation}
This establishes the relationship. The alternative Stratonovich QSDE where
the noise coefficients lie to the right of the noise is handled with (3.12).

In (ii), the process\ $A_{t}^{11}+A_{t}^{10}+A_{t}^{10}+A_{t}^{00}$, in the
Fock vacuum state, corresponds to the Poisson process. Here $%
X_{t}^{11}=X_{t}^{10}=X_{t}^{10}=X_{t}^{00}-\tilde{v}(X_{t},t)=\tilde{\mu}%
\left( X_{t},t\right) $. To obtain the Stratonovich QSDE with integrators on
the right, the terms leading to differences are those in $dA^{11}$ and $%
dA^{10}$; from (3.11) again one has 
\begin{eqnarray}
\overline{f\left( X_{t}\right) }\left\{ dA_{t}^{11}+A_{t}^{10}\right\} &=&%
\frac{1}{2}\left\{ f\left( X_{t}+\tilde{\mu}\right) +f\left( X_{t}\right)
\right\} \left\{ dA_{t}^{11}+A_{t}^{10}\right\}  \TCItag{3.15} \\
&&+\frac{1}{2}\sum_{n}f_{n}\sum_{p_{1},p_{2}}^{n}\left( X_{t}\right) ^{p_{1}}%
\tilde{\mu}\left( X_{t}+\tilde{\mu}\right) ^{p_{2}}\left\{
dA_{t}^{01}+A_{t}^{00}\right\} .  \notag
\end{eqnarray}
The coefficient of the last term sums to $\frac{1}{2}\left\{ f\left( X_{t}+%
\tilde{\mu}\left( X_{t},t\right) \right) -f\left( X_{t}\right) \right\} $,
and so one concludes 
\begin{eqnarray}
&&\overline{f\left( X_{t}\right) }\left\{
dA_{t}^{11}+dA_{t}^{10}+dA_{t}^{10}+dA_{t}^{00}\right\}  \notag \\
&&+\frac{1}{2}\left\{ f\left( X_{t}+\tilde{\mu}\left( X_{t},t\right) \right)
-f\left( X_{t}\right) \right\} \left\{
dA_{t}^{11}+dA_{t}^{10}+dA_{t}^{10}+dA_{t}^{00}\right\} .  \label{3.16}
\end{eqnarray}
\end{proof}

\bigskip

\section{Quantum white noise representation}

There is a more natural way to look at this problem. Take as admissible
space the subset $\mathcal{S}\subset L^{2}\left( \mathbb{R}^{+},\mathcal{K}%
\right) $ got by taking the sup-norm completion of the square-integrable
step functions on $\mathbb{R}^{+}$. That is, S is the set of
square-integrable regulated functions and for $f\in \mathcal{S}$ one is
guaranteed that the past and future instant limits $f(t^{-})$ and $f(t^{+})$
exist at each $t>0$, cf. Dieudonne [21]. Note that the projection
requirement of admissible spaces rules out the Schwartz functions; the space
of test functions $\mathcal{S}$ is in a sense the most natural choice of it
as the widest space on which integral approximations can be based and also
contains the functions of bounded variation which are the natural space on
which to discuss functional integral transforms. One defines functionals $%
\delta _{\pm }\left( t\right) $ and $\delta _{\ast }\left( t\right) $ on $%
\mathcal{S}$ by 
\begin{equation}
\langle \delta _{\pm }\left( t\right) ,f\rangle =f\left( t^{\pm }\right)
,\quad \langle \delta _{\ast }\left( t\right) ,f\rangle =\frac{1}{2}%
\{f\left( t^{+}\right) +f\left( t^{-}\right) \}.  \tag{4.1}
\end{equation}
That is, the action of $\delta _{\ast }\left( t\right) $ on a function of a
time variable is to give the average of the immediate past and future values
at time $t$. The action of these functionals can be extended by linearity
from $\mathcal{S}$ to $\mathcal{S}^{\prime }$ if the following
identifications are made: 
\begin{eqnarray}
\langle \delta _{-}\left( t\right) ,\delta _{+}\left( s\right) \rangle
&:&=\delta _{+}\left( t-s\right) ,  \notag \\
\langle \delta _{+}\left( t\right) ,\delta _{-}\left( s\right) \rangle
&:&=\delta _{-}\left( t-s\right) ,  \notag \\
\langle \delta _{+}\left( t\right) ,\delta _{+}\left( s\right) \rangle
&:&=\delta _{\ast }\left( t-s\right) ,  \notag \\
\langle \delta _{-}\left( t\right) ,\delta _{-}\left( s\right) \rangle
&:&=\delta _{\ast }\left( t-s\right) ,  \TCItag{4.2}
\end{eqnarray}

\noindent \bigskip \textbf{DEFINITION 4.1.} A \emph{basic step function} on $%
\mathbb{R}^{n}$ is the characteristic function of a set of the form $\left\{
\left( t_{1},\cdots ,t_{n}\right) :0\leq t_{j\left( 1\right) }-c_{1}\leq
\cdots \leq t_{j\left( r\right) }-c_{r}\leq c_{0}\right\} $, where $1\leq
r\leq n$, the $j(l),...,j(r)$ are distinct elements of $\left\{
1,...,n\right\} $, and $c_{0},c_{1},\cdots ,c_{r}$ are constants, and also
where any of the strict inequalities can be replaced by ordinary ones. A 
\emph{simple function} on $\mathbb{R}^{n}$ is a finite linear combination of
such step functions. The space of multi-dimensional regulated functions $%
\mathcal{S}_{n}$, is taken to be the sup-norm completion of these simple
functions.

For $f\in \mathcal{S}$, the individual limits of\ $f\left( t_{1},\cdots
,t_{n}\right) $ exist as the $t_{j}\rightarrow a_{j}^{\pm }$ under the
conditions $t_{\sigma \left( 1\right) }<\cdots <t_{\sigma \left( n\right) }$
for all $a_{j}\in \mathbb{R}$, and for each permutation $\sigma $ of $%
\left\{ 1,\cdots ,n\right\} $.

\bigskip

\noindent \textbf{DEFINITION 4.2.} \emph{Quantum white noises} are defined
on EXP($\mathcal{S}$) by $a_{\pm }^{\sharp }\left( t\right) :=A^{\sharp
}\left( \delta _{\pm }\left( t\right) \right) $ and $a_{\ast }^{\sharp
}\left( t\right) :=A^{\sharp }\left( \delta _{\ast }\left( t\right) \right) $%
. Explicitly, for $f\in \mathcal{S}$, one has $a_{\pm }\left( t\right) \Psi
\left( f\right) =f\left( t^{\pm }\right) \Psi \left( f\right) $, etc.

From (4.2), the nontrivial commutations between the $a_{\pm }^{\sharp
}\left( t\right) $ are given by 
\begin{eqnarray}
\lbrack a_{-}\left( t\right) ,a_{+}^{\dag }\left( s\right) ] &=&\delta
_{+}\left( t-s\right) ,  \notag \\
\lbrack a_{+}\left( t\right) ,a_{-}^{\dag }\left( s\right) ] &=&\delta
_{-}\left( t-s\right) ,  \notag \\
\lbrack a_{+}\left( t\right) ,a_{+}^{\dag }\left( s\right) ] &=&\delta
_{\ast }\left( t-s\right) ,  \notag \\
\lbrack a_{-}\left( t\right) ,a_{-}^{\dag }\left( s\right) ] &=&\delta
_{\ast }\left( t-s\right) ,  \TCItag{4.3}
\end{eqnarray}

The linearity of the extended functionals then implies that the pair of
processes $\left\{ a_{\ast }^{\sharp }\left( t\right) ;t>0\right\} $ satisfy
the following canonical commutation relations (CCR) 
\begin{equation}
\lbrack a_{\ast }\left( t\right) ,a_{\ast }\left( s\right) ]=0=[a_{\ast
}^{\dag }\left( t\right) ,a_{\ast }^{\dag }\left( s\right) ],\quad \lbrack
a_{\ast }\left( t\right) ,a_{\ast }^{\dag }\left( s\right) ]=\delta _{\ast
}\left( t-s\right) .  \tag{4.4}
\end{equation}
One further has the following functional distribution 
\begin{equation}
\left\langle \exp \left( \int {}_{0}^{\infty }dt\left\{ f\left( t\right)
a_{\ast }^{\dag }\left( t\right) +f\left( t\right) ^{\ast }a_{\ast }^{\dag
}\left( t\right) \right\} \right) \right\rangle =\exp \left[ -\frac{1}{2}%
\left\| f\right\| ^{2}\right] ,  \tag{4.5}
\end{equation}
where $f\in \mathcal{S}$, and $\left\| f\right\| ^{2}=\int_{0}^{\infty
}\left| f\left( t\right) \right| ^{2}dt$, and the expectation is in the Fock
vacuum state $\Psi \left( 0\right) $. The key feature of (4.4) is that,
along with the specification of the state (4.5), it contains all information
concerning the chaotic expansions. Any integral of the form $\int_{\mathbb{R}%
^{n}}dt_{1}\cdots dt_{n}\varphi \left( t_{1},\cdots ,t_{n}\right) a_{\ast
}^{\sharp \left( 1\right) }\left( t_{1}\right) \cdots a_{\ast }^{\sharp
\left( n\right) }\left( t_{n}\right) $, for $\varphi \in \mathcal{S}$, can
be evaluated and, in particular, one may take $\varphi $ to be simplicial.

The following connection [10,17] exists between the quantum stochastic
calculus and white noise calculus.

\noindent \textbf{THEOREM 4.3.} Let $(X_{t})_{t\geq 0}$ be the solution of
the QSDE $dX_{t}=X_{t}^{\alpha \beta }\otimes dA_{t}^{\alpha \beta }$ with
the $\left( X_{t}^{\alpha \beta }\right) _{t\geq 0}$ adapted processes based
on $(\mathcal{D},\mathcal{S})$, then the QSDE can be represented as 
\begin{eqnarray*}
dX_{t} &=&\left\{ a_{\ast }^{\dag }\left( t\right) X_{t}^{11}a_{\ast }\left(
t\right) +a_{\ast }^{\dag }\left( t\right) X_{t}^{10}+X_{t}^{01}a_{\ast
}\left( t\right) +X_{t}^{00}\right\} dt \\
&=&\left[ 1,a_{\ast }^{\dag }\left( t\right) \right] \left[ 
\begin{array}{cc}
X_{t}^{00} & X_{t}^{01} \\ 
X_{t}^{10} & X_{t}^{11}
\end{array}
\right] \left[ 
\begin{array}{c}
1 \\ 
a_{\ast }\left( t\right)
\end{array}
\right] .
\end{eqnarray*}

\noindent \textbf{Remarks}

i) The Wiener and Poisson processes are represented by $B_{t}=\int_{0}^{t}%
\left( a_{\ast }^{\dag }\left( s\right) +a_{\ast }\left( s\right) \right) ds$%
, $N_{t}=\int_{0}^{t}\left( 1+a_{\ast }\left( s\right) \right) ^{\dag
}\left( 1+a_{\ast }\left( s\right) \right) ds$, respectively. Their chaotic
expansions can readily be obtained.

ii) A Stratonovich QSDE is an equation of the form 
\begin{eqnarray}
dX_{t} &=&\left\{ a_{\ast }^{\dag }\left( t\right) a_{\ast }\left( t\right) 
\overline{E_{t}}^{R}+a_{\ast }^{\dag }\left( t\right) \overline{F_{t}}%
^{R}+a_{\ast }\left( t\right) \overline{G_{t}}^{R}+\overline{H_{t}}%
^{R}\right\} dt  \TCItag{4.7a} \\
dX_{t} &=&\left\{ \overline{E_{t}}^{L}a_{\ast }^{\dag }\left( t\right)
a_{\ast }\left( t\right) +\overline{F_{t}}^{L}a_{\ast }^{\dag }\left(
t\right) +\overline{G_{t}}^{L}a_{\ast }\left( t\right) +\overline{H_{t}}%
^{L}\right\} dt  \TCItag{4.7b}
\end{eqnarray}
Equation (4.7a) is the left handed version, and (4.7b) is the right handed
version.

If the conserver terms are ignored, then one sees that the quantum Ito
calculus corresponds to the Wick ordering scheme while the quantum
Stratonovich corresponds to the Weyl scheme. This point of view can be of
help in understanding the related distinctions which arise in the theory of
phase space path integrals [22]. The anti-Wick ordering scheme gives
time-reversed quantum Brownian motion.

iii) The classical Wiener integral is then represented as

\begin{equation}
X_{t}=X_{0}+\int_{0}^{t}\left\{ \overline{v\left( X_{s},s\right) }+\left[
a_{\ast }^{\dag }\left( s\right) +a_{\ast }\left( s\right) \right] \overline{%
\sigma \left( X_{s},s\right) }\right\} ds  \tag{4.8a}
\end{equation}
or 
\begin{equation*}
X_{t}=X_{0}+\int_{0}^{t}\left\{ \underline{\tilde{v}\left( X_{s},s\right) }%
+a_{\ast }^{\dag }\left( s\right) \underline{\tilde{\sigma}\left(
X_{s},s\right) }+\underline{\tilde{\sigma}\left( X_{s},s\right) }a_{\ast
}\left( s\right) \right\} ds
\end{equation*}
where $v,\sigma $ and $\tilde{v},\tilde{\sigma}$ are related by (2.11).

Likewise the classical Poissonian integral is represented as 
\begin{equation}
X_{t}=X_{0}+\int_{0}^{t}\left\{ \overline{v\left( X_{s},s\right) }+\left[
1+a_{\ast }\left( s\right) \right] ^{\dag }\left[ 1+a_{\ast }\left( s\right) %
\right] \overline{\mu \left( X_{s},s\right) }\right\} ds  \label{4.9a}
\end{equation}
or 
\begin{equation*}
X_{t}=X_{0}+\int_{0}^{t}\left\{ \underline{\tilde{v}\left( X_{s},s\right) }+%
\left[ 1+a_{\ast }\left( s\right) \right] ^{\dag }\underline{\tilde{\mu}%
\left( X_{s},s\right) }\left[ 1+a_{\ast }\left( s\right) \right] \right\} ds
\end{equation*}
where $v,\mu $ and $\tilde{v},\tilde{\mu}$ are related by (2.12).

iv) Let $X_{t}=x_{0}+\int_{0}^{t}X_{t}^{\alpha \beta }\otimes dA_{t}^{\alpha
\beta }$, $Y_{t}=$\ $y_{0}+\int_{0}^{t}Y_{t}^{\alpha \beta }\otimes
dA_{t}^{\alpha \beta }$ be processes based on $(\mathcal{D},\mathcal{S})$,
then 
\begin{eqnarray*}
\int_{0}^{t}\overline{X_{s}}dY_{s} &=&\int_{0}^{t}ds\left(
x_{0}+\int_{0}^{s}ds\left\{ a_{\ast }^{\dag }\left( s\right)
X_{s}^{11}a_{\ast }\left( s\right) +a_{\ast }^{\dag }\left( s\right)
X_{s}^{10}+X_{s}^{01}a_{\ast }\left( s\right) +X_{s}^{00}\right\} \right) \\
&&\times \left\{ a_{\ast }^{\dag }\left( s\right) Y_{s}^{11}a_{\ast }\left(
s\right) +a_{\ast }^{\dag }\left( s\right) Y_{s}^{10}+Y_{s}^{01}a_{\ast
}\left( s\right) +Y_{s}^{00}\right\} \\
&=&\int_{0}^{t}ds\left\{ a_{\ast }^{\dag }\left( s\right)
X_{s}Y_{s}^{11}a_{\ast }\left( s\right) +a_{\ast }^{\dag }\left( s\right)
X_{s}Y_{s}^{10}+X_{s}Y_{s}^{01}a_{\ast }\left( s\right)
+X_{s}Y_{s}^{00}\right\} \\
&&+\int_{0}^{t}ds\int_{0}^{s}du\delta _{\ast }\left( u-s\right) [a_{\ast
}^{\dag }\left( u\right) X_{u}^{11}Y_{s}^{11}a_{\ast }\left( s\right) \\
&&+a_{\ast }^{\dag }\left( u\right)
X_{u}^{11}Y_{s}^{10}+X_{u}^{01}Y_{s}^{11}a_{\ast }\left( s\right)
+X_{u}^{01}Y_{s}^{10}] \\
&=&\int_{0}^{t}X_{s}dY_{s}+\frac{1}{2}\int_{0}^{t}ds\left[
X_{s}^{01}+a_{\ast }^{\dag }\left( s\right) X_{s}^{11}\right] \left[
Y_{s}^{10}+Y_{s}^{11}a_{\ast }\left( s\right) \right] .
\end{eqnarray*}

That is, $\int_{0}^{t}\overline{X_{s}}dY_{s}=\int_{0}^{t}X_{s}dY_{s}+\frac{1%
}{2}\int_{0}^{t}dX_{s}dY_{s}$.

Thus the algebraic product $X_{t}Y_{t}$ of the white noise representations
gives the correct product as quantum stochastic processes.

v) From the above considerations, it follows that the following formal
manipulations are allowed 
\begin{eqnarray*}
\int_{0}^{T}X_{t}\otimes dA_{t}^{01}-\int_{0}^{T}dA_{t}^{01}\overline{X_{t}}
&=&\int_{0}^{T}\left[ X_{t},a_{\ast }\left( t\right) \right] dt \\
&=&\int_{0}^{T}dt\int_{0}^{t}ds\left[ a_{\ast }^{\dag }\left( t\right)
X_{t}^{11}a_{\ast }\left( t\right) +a_{\ast }^{\dag }\left( t\right)
X_{t}^{10}+X_{t}^{01}a_{\ast }\left( t\right) +X_{t}^{00},a_{\ast }\left(
s\right) \right] \\
&=&-\int_{0}^{T}dt\int_{0}^{t}ds\delta _{\ast }\left( t-s\right) \left[
X_{s}^{11}a_{\ast }\left( s\right) +X_{s}^{10}\right] \\
&=&-\frac{1}{2}\int_{0}^{T}dt\left[ X_{t}^{11}a_{\ast }\left( t\right)
+X_{t}^{10}\right] \\
&=&-\frac{1}{2}\int_{0}^{T}dt\left[ X_{t}^{11}\otimes
dA_{t}^{10}+X_{t}^{10}\otimes dA_{t}^{00}\right] .
\end{eqnarray*}
More generally, for polynomial $f$, 
\begin{equation*}
df\left( X_{t}\right) =\left\{ a_{\ast }^{\dag }\left( t\right)
f(X_{t})^{11}a_{\ast }\left( t\right) +a_{\ast }^{\dag }\left( t\right)
f(X_{t})^{10}+f(X_{t})^{01}a_{\ast }\left( t\right) +f(X_{t})^{00}\right\}
dt,
\end{equation*}
and formal manipulation leads to 
\begin{equation*}
\left[ f\left( X_{t}\right) ,a_{\ast }\left( t\right) \right] =-\frac{1}{2}%
f(X_{t})^{11}a_{\ast }\left( t\right) -\frac{1}{2}f(X_{t})^{10},
\end{equation*}
or under rearrangement 
\begin{equation*}
\frac{1}{2}\underline{f\left( X_{t}+X_{t}^{11}\right) +f\left( X_{t}\right) }%
a_{\ast }\left( t\right) =a_{\ast }\left( t\right) \overline{f\left(
X_{t}\right) }-\frac{1}{2}f\left( X_{t}\right) ^{10}.
\end{equation*}
This is the same as the third relation of (3.12).

\section{Asymptotic quantum stochastic limits; convergence ansatz}

Let $\mathcal{K}_{0}$ be a fixed Hilbert space and $S(t)=e^{i\Omega t}$ be a
strongly continuous one-parameter unitary group on $\mathcal{K}_{0}$. If $%
\mathcal{\tilde{K}}_{0}$ is the subspace of $\mathcal{K}_{0}$ such that $%
\int_{-\infty }^{\infty }dt\left| \langle S\left( t\right) g,f\rangle
\right| <\infty $ whenever $f,g\in \mathcal{K}_{0}$, then introducing the
sesquilinear form 
\begin{equation}
\gamma \left( f,g\right) =\int_{-\infty }^{\infty }dt\langle S\left(
t\right) g,f\rangle  \tag{5.1}
\end{equation}
the Hilbert space completion of $\mathcal{\tilde{K}}_{0}$, with subspace $%
\{k:\gamma (k,k)=0\}$ factored out, will be denoted by $\mathcal{K}$. The
inner product on $\mathcal{K}$ will be taken to be that inherited from\ $%
\mathcal{K}_{0}$ which will be denoted $\langle \cdot ,\cdot \rangle _{%
\mathcal{K}}$. For each $g\in \mathcal{K}$ the mapping $t\rightarrow S\left(
t\right) g$ is Bochner integrable. The following operator is well defined on 
$\mathcal{K}$: 
\begin{equation}
S_{+}:=\int_{0}^{\infty }S\left( t\right) dt\equiv \frac{i}{\Omega +i0^{+}}%
=\pi \delta \left( \Omega \right) +iPV\frac{1}{\Omega }  \tag{5.2}
\label{5.2}
\end{equation}
and the notation $S_{-}:=(S_{+})^{\dag },\Gamma =2ReS_{+}$ and $\Sigma
:=ImS_{+}$ shall be adopted. Noting that $\Gamma \geq 0$ and that $[\Gamma
,\Sigma ]=0$, it follows that there exists $Z=X+iY$ with $X$ and $Y$
self-adjoint on $\mathcal{K}$ such that 
\begin{equation}
X^{2}=\frac{1}{4}\Gamma ,\quad XY+YX=-\Sigma .  \tag{5.3}
\end{equation}

For each $\gamma >0$, one considers $\mathcal{H}_{\lambda }$ a copy of $%
\Gamma _{+}\left( \mathcal{K}\right) $ with Fock vacuum vector denoted as $%
\Psi _{\lambda }$ and $A_{\lambda }^{\sharp }\left( \cdot \right) $ the
creator/destroyer maps. For $g\in \mathcal{K},t>0$, one introduces the
following operators on $\mathcal{H}_{\lambda }$%
\begin{equation}
a_{\lambda }^{\sharp }\left( t,g\right) =A_{\lambda }^{\sharp }\left( \frac{1%
}{\lambda }S\left( t/\lambda ^{2}\right) g\right) .  \tag{5.4}
\end{equation}
From the CCR one has 
\begin{equation}
\left[ a_{\lambda }\left( t,g\right) ,a_{\lambda }^{\dag }\left( s,f\right) %
\right] =\frac{1}{\lambda ^{2}}\langle S\left( \frac{t-s}{\lambda ^{2}}%
\right) g,f\rangle _{\mathcal{K}}  \tag{5.5}
\end{equation}
which says roughly that these operators have auto-correlation time of the
order $\lambda ^{2}$. As $\lambda \rightarrow 0$, one expects these
operators to become white noises, however, taking account of the previous
section, the functional limit of (5.5) can be interpreted more concisely as 
\begin{equation}
\left[ a\left( t,g\right) ,a^{\dag }\left( s,f\right) \right] =\langle
S_{+}g,f\rangle _{\mathcal{K}}\delta _{+}\left( t-s\right) +\langle
S_{-}g,f\rangle _{\mathcal{K}}\delta _{-}\left( t-s\right)  \tag{5.6}
\end{equation}
The limit operators $a^{\sharp }\left( t,g\right) $ are interpreted as the
operators defined on $\mathcal{H}:=\Gamma _{+}\left( L^{2}\left( \mathbb{R}%
\right) \underline{\otimes }\mathcal{K}\right) $ by 
\begin{equation}
a\left( t,g\right) \Psi \left( \phi \underline{\otimes }f\right) :=\langle
\delta _{+}\left( t\right) \underline{\otimes }Zg+\delta _{-}\left( t\right) 
\underline{\otimes }Z^{\dag }g,\phi \underline{\otimes }f\rangle \;\Psi
\left( \phi \underline{\otimes }f\right)  \tag{5.7}
\end{equation}
with $a^{\dag }(t,g)$ the adjoint of $a(t,g)$.

\noindent \textbf{THEOREM 5.1.} The processes $\{a_{\lambda }^{\sharp
}(t,g):t\geq 0,g\in \mathcal{K}\}$ on $\mathcal{H}_{\lambda }$ with the
state $\Psi _{\lambda }$ converge in Fock vacuum expectation as $\lambda
\rightarrow 0$ to the quantum white noises $\{a^{\sharp }(t,g):t\geq 0,g\in 
\mathcal{K}\}$ on $\mathcal{H}$ with the state $\Psi $. That is, for each
integer $n\geq 0$, the following limits hold on $\mathcal{S}_{n}^{^{\prime }}
$, 
\begin{equation}
\lim_{\lambda \rightarrow 0^{+}}\langle \Psi _{\lambda },a_{\lambda
}^{\sharp \left( 1\right) }\left( t_{1},g_{1}\right) \cdots a_{\lambda
}^{\sharp \left( n\right) }\left( t_{n},g_{n}\right) \Psi _{\lambda }\rangle
_{\mathcal{H}_{\lambda }}=\langle \Psi ,a^{\sharp \left( 1\right) }\left(
t_{1},g_{1}\right) \cdots a^{\sharp \left( n\right) }\left(
t_{n},g_{n}\right) \Psi \rangle _{\mathcal{H}},  \tag{5.8}
\end{equation}
for all $t_{1},\cdots ,t_{n}\geq 0$, $g_{1},...,g_{n}\in \mathcal{K}$ and
choices of creators and/or destroyers.

\begin{proof}
From the Gaussianity of the pre-limit processes it suffices to consider the
two-point functions. For $\phi \in S_{2}$, let 
\begin{eqnarray*}
I_{\lambda } &=&\int_{\mathbb{R}^{+}\times \mathbb{R}^{+}}dsdt\;\phi \left(
t,s\right) \frac{1}{\lambda ^{2}}\langle S\left( \frac{t-s}{\lambda ^{2}}%
\right) g,f\rangle _{\mathcal{K}} \\
&=&\int_{0}^{\infty }du\int_{-u/\lambda ^{2}}^{u/\lambda ^{2}}d\tau \,\phi
\left( u+\lambda ^{2}\tau ,u-\lambda ^{2}\tau \right) \langle S\left( \tau
\right) g,f\rangle _{\mathcal{K}},
\end{eqnarray*}
where the change of variables $u=t+s,\tau =\left( t-s\right) /\lambda ^{2}$
was made. If $J_{\theta }$ is the expression obtained from $I_{\lambda }$,
by replacing the $t$-limits of integration by $\pm \infty $, then $\left|
I_{\lambda }-J_{\lambda }\right| \rightarrow 0$ as $\lambda \rightarrow
0^{+} $ uniformly since $f,g\in \mathcal{K}$. Moreover, since $\phi $ is $%
L^{2}$-regulated, $J_{\lambda }$ converges uniformly to 
\begin{equation*}
\int_{0}^{\infty }du\,\left\{ \phi \left( u^{+},u^{-}\right) \langle
S_{+}g,f\rangle _{\mathcal{K}}+\phi \left( u^{-},u^{+}\right) \langle
S_{-}g,f\rangle _{\mathcal{K}}\right\}
\end{equation*}
which establishes (5.6).

Next the CCR of the limit noises must be checked: 
\begin{eqnarray*}
\left[ a\left( t,g\right) ,a^{\dag }\left( s,f\right) \right] &=&\langle
\delta _{+}\left( t\right) \underline{\otimes }Zg+\delta _{-}\left( t\right) 
\underline{\otimes }Z^{\dag }g,\delta _{+}\left( s\right) \underline{\otimes 
}Zf+\delta _{-}\left( s\right) \underline{\otimes }Z^{\dag }f\rangle _{%
\mathcal{H}} \\
&=&\langle (Z^{\dag })^{2}+\frac{1}{2}(Z^{\dag }Z+ZZ^{\dag })g,f\rangle _{%
\mathcal{K}}\delta _{+}\left( t-s\right) \\
&&+\langle (Z^{2}+\frac{1}{2}(Z^{\dag }Z+ZZ^{\dag })g,f\rangle _{\mathcal{K}%
}\delta _{-}\left( t-s\right) ,
\end{eqnarray*}
but $(Z^{\dag })^{2}+\frac{1}{2}(Z^{\dag }Z+ZZ^{\dag })=2X^{2}-i(XY+YX)=%
\frac{1}{2}\Gamma +i\Sigma =S_{+}$ and so (5.6) is recovered.
\end{proof}

\bigskip

\noindent \textbf{DEFINITION 5.2.} Let $\Psi _{\lambda }\left( h\right) $
denote the exponential vector on $\mathcal{H}_{\lambda }=\Gamma _{+}\left( 
\mathcal{K}\right) $ for $k\in \mathcal{K}$. Collective exponential vectors
are defined, for $f\in \mathcal{S}$, by 
\begin{equation}
\Psi \left( f,k,\lambda \right) :=\Psi _{\lambda }\left( \int_{0}^{\infty
}dt\,f\left( t\right) S\left( t/\lambda ^{2}\right) k\right) .  \tag{5.9}
\end{equation}

The set of such collective exponential vectors is denoted $EXP_{\lambda }(%
\mathcal{S},\mathcal{K})$.

Noting that $\left\langle \Psi \left( f,k,\lambda \right) ,\Psi \left(
f^{\prime },k^{\prime },\lambda \right) \right\rangle _{\mathcal{H}_{\lambda
}}\rightarrow \exp \left\{ \gamma \left( k,k^{\prime }\right) \langle
f,f^{\prime }\rangle _{L^{2}\left( \mathbb{R}^{+}\right) }\right\} $ as $%
\lambda \rightarrow 0^{+}$, \ it is natural to associate the limit $\Psi
\left( f\underline{\otimes }\Gamma ^{\frac{1}{2}}k\right) $ in $\mathcal{H}$
with $\Psi \left( f,k,\lambda \right) $ by virtue that $\gamma \left(
k,k\right) =\langle \Gamma k,k\rangle _{\mathcal{K}}=\left\| \Gamma
^{1/2}k\right\| _{\mathcal{K}}^{2}$.

In the following we shall understand all processes to be over a common
domain $\mathcal{D}$ of $\mathcal{H}_{0}$, as outlined in Section 3. A
family of operators $\left( Z_{t}\left( \lambda \right) \right) $ on \ $%
\mathcal{H}_{0}\underline{\otimes }\mathcal{H}_{\lambda }$ is said to
converge to a process $Z_{t}$ on $\mathcal{H}_{0}\underline{\otimes }%
\mathcal{H}$ as $\lambda \rightarrow 0^{+}$ weakly in matrix elements [9] if
for all $n,t_{1},\cdots ,t_{n}$, and for all $u,u^{\prime }\in \mathcal{D}$, 
$f,f^{\prime }\in \mathcal{S}$ and $k,k^{\prime }\in \mathcal{K}$ one has 
\begin{equation*}
\lim_{\lambda \rightarrow 0^{+}}\left\langle u\otimes \Psi \left(
f,k,\lambda \right) ,Z_{t_{i}}\left( \lambda \right) \cdots Z_{t_{n}}\left(
\lambda \right) u^{\prime }\otimes \Psi \left( f^{\prime },k^{\prime
},\lambda \right) \right\rangle =\left\langle u\otimes \Psi \left( f%
\underline{\otimes }k\right) ,Z_{t_{i}}\left( \lambda \right) \cdots
Z_{t_{n}}\left( \lambda \right) u^{\prime }\otimes \Psi \left( f^{\prime }%
\underline{\otimes }k^{\prime }\right) \right\rangle .
\end{equation*}

Typically one would also like $Z_{t}\left( \lambda \right) $ to converge to
an adapted process and for this reason the next definition is formulated.

\noindent \textbf{DEFINITION 5.3.} A process $\left( Z_{t}\left( \lambda
\right) \right) _{t\geq 0}$ on $\mathcal{H}_{0}\underline{\otimes }\mathcal{H%
}$ is said to be adaptable if, for each $t\geq 0$ and for all $s>t$, $g\in 
\mathcal{K}$, one has 
\begin{equation}
\left[ Z_{t}\left( \lambda \right) ,a_{\lambda }^{\sharp }\left( s,g\right) %
\right] =O_{t,s}\left( \lambda \right) ,  \tag{5.10}
\end{equation}
whereby a process $W=W_{t,s}(h)$ is said to be $O_{t,s}\left( \lambda
\right) $ if $\int_{0}^{T}ds\int_{0}^{s}dtW_{t,s}\left( \lambda \right) $%
vanishes in matrix element limits. Further one requires that, for all $u\in 
\mathcal{H}_{0}$, $f\in \mathcal{S}$\ and \ $g\in \mathcal{K}$, and $\lambda 
$ in a neighborhood of $0^{+}$, 
\begin{equation}
\int_{0}^{t}ds\,\left\| Z_{s}\left( \lambda \right) \,u\otimes \Psi \left(
f,k,\lambda \right) \right\| ^{2}<C,  \tag{5.11}
\end{equation}
where $C$ depends at most on $u,f$ and $g$.

Let $X_{t}^{\alpha \beta }\left( \lambda \right) $ be operators on $\mathcal{%
H}_{0}\underline{\otimes }\mathcal{H}$ which depend on $a_{\lambda }^{\sharp
}\left( t,\cdot \right) $ for $s<t$. Note that, since the $a_{\lambda
}^{\sharp }\left( s,g\right) $ have finite auto-correlation time, it does
not follow that they will commute with $X_{t}^{\alpha \beta }\left( \lambda
\right) $ for $s>t$, however the assumption shall be made that they are
adaptable. Consider then ODES of the type 
\begin{eqnarray}
\frac{d}{dt}X_{t}\left( \lambda \right) &=&a_{\lambda }^{\dag }\left(
t,g_{1}\right) X_{t}^{11}\left( \lambda \right) a_{\lambda }\left(
t,g_{2}\right) +a_{\lambda }^{\dag }\left( t,g_{3}\right) X_{t}^{10}\left(
\lambda \right) +X_{t}^{01}\left( \lambda \right) a_{\lambda }\left(
t,g_{4}\right) +X_{t}^{00}\left( \lambda \right) ,  \TCItag{5.12} \\
X_{0}\left( \lambda \right) &=&x_{0}\in \mathcal{B}\left( \mathcal{H}%
_{0}\right) .  \notag
\end{eqnarray}
In this form the ODE has creators and destroyers in normal ordered from. The
objective of the remainder of this section is to show that (5.12) converges
to a well-defined QSDE under the ansatz that its coefficients are adaptable.
For convenience, the K-state will be a fixed $g$ and it will be supposed
that $\langle S_{\pm }g,g\rangle _{\mathcal{K}}=\frac{1}{2}$ for the rest of
this section. The $g$-dependence in the pre-limit and limit noises and
collective exponential vectors will be dropped, and one notes that in this
case the limit noise is just the fields $a_{\ast }\left( t\right) $
introduced in the previous section.

\noindent \textbf{THEOREM 5.4.} Suppose $X_{t}(\lambda )$ is the unique
solution to a Wick ordered ODE with adaptable coeficients $X_{t}^{\alpha
\beta }\left( \lambda \right) $) converging to processes $X_{t}^{\alpha
\beta }$ so that the QSDE $dX_{t}=X_{t}^{\alpha \beta }\otimes
dA_{t}^{\alpha \beta }$, $X_{0}=x_{0}$, has again unique solution $X_{t}$.
Then $X_{t}\left( \lambda \right) $ converges to $X_{t}$, weakly in matrix
elements as $\lambda \rightarrow 0^{+}$.

\begin{proof}
Part (i). Note that one has now taken $\gamma (g,g)=1$ and writes $\Psi
\left( f\right) $ for $\Psi \left( f\underline{\otimes }\Gamma ^{\frac{1}{2}%
}g\right) $. Let $\varphi _{1},\varphi _{2}\in \mathcal{H}_{0}$, $f,k\in
L^{2}(\mathbb{R}^{+})$, then as $\lambda \rightarrow 0$, 
\begin{eqnarray*}
&&\langle \varphi _{1}\underline{\otimes }\Psi \left( f,\lambda \right)
,\left\{ X_{t}\left( \lambda \right) -x_{0}\right\} \varphi _{2}\underline{%
\otimes }\Psi \left( k,\lambda \right) \rangle \\
&=&\int_{0}^{t}ds\langle \varphi _{1},\left\{ \int_{0}^{\infty }du\frac{%
f\left( u\right) ^{\ast }}{\lambda ^{2}}\langle g,S\left( \frac{u-s}{\lambda
^{2}}\right) g\rangle X_{s}^{11}\left( \lambda \right) \int_{0}^{\infty }dv%
\frac{k\left( v\right) }{\lambda ^{2}}\langle g,S\left( \frac{v-s}{\lambda
^{2}}\right) g\rangle +\cdots \right\} \varphi _{2}\rangle _{\mathcal{H}_{0}}
\\
&\rightarrow &\int_{0}^{t}ds\langle \varphi _{1},\left\{ f\left( s\right)
^{\ast }X_{s}^{11}k\left( s\right) +f\left( s\right) ^{\ast
}X_{s}^{10}+X_{s}^{01}k\left( s\right) +X_{s}^{00}\right\} \varphi
_{2}\rangle _{\mathcal{H}_{0}} \\
&=&\langle \varphi _{1}\underline{\otimes }\Psi \left( f\right)
,\int_{0}^{t}X_{s}^{\alpha \beta }\otimes dA_{s}^{\alpha \beta }\,\varphi
_{2}\underline{\otimes }\Psi \left( k\right) \rangle _{\mathcal{H}_{0}%
\underline{\otimes }\mathcal{H}}.
\end{eqnarray*}
This amounts to saying that the approximation holds at the level of the
QSDE, that is, $X_{t}\left( \lambda \right) $ converges to $X_{t}$ in first
moment.

Part (ii). It is easy to see that higher moments $X_{t_{1}}\left( \lambda
\right) ,\cdots ,X_{t_{n}}\left( \lambda \right) $ decouple in collective
coherent states, under the assumptions of the theorem, provided the $t_{j}$
are distinct. To deal with the case where several indices are equal, one
actually shows that such moments converge to the appropriate equal time Ito
products. This is established by first showing that if $Y_{t}\left( h\right) 
$ is a similarly described process, then the product $Y_{t}\left( h\right)
X_{t}\left( h\right) $\ converges to $X_{t}Y_{t}$ in QSDE.

Multiplying the pre-limit operators, one obtains 
\begin{eqnarray*}
&&\left( X_{t}\left( \lambda \right) -x_{0}\right) \left( Y_{t}\left(
\lambda \right) -y_{0}\right) \\
&=&\int_{0}^{t}du\int_{0}^{t}dv\left\{ a_{\lambda }^{\dag }\left( u\right)
X_{u}^{11}\left( \lambda \right) a_{\lambda }\left( u\right) +\cdots
\right\} \left\{ a_{\lambda }^{\dag }\left( v\right) YX_{v}^{11}\left(
\lambda \right) a_{\lambda }\left( v\right) +\cdots \right\} ,
\end{eqnarray*}
there are 4 x 4 separate integrals here, examination of the first will be
sufficient to determine the general pattern; one has 
\begin{eqnarray*}
&&\int_{0}^{t}du\int_{0}^{t}dv\,a_{\lambda }^{\dag }\left( u\right)
X_{u}^{11}\left( \lambda \right) a_{\lambda }\left( u\right) a_{\lambda
}^{\dag }\left( v\right) Y_{v}^{11}\left( \lambda \right) a_{\lambda }\left(
v\right) \\
&=&\int_{0}^{t}du\int_{0}^{t}dv\,a_{\lambda }^{\dag }\left( u\right)
X_{u}^{11}\left( \lambda \right) \left\{ a_{\lambda }^{\dag }\left( v\right)
a_{\lambda }\left( u\right) +\frac{1}{\lambda ^{2}}\langle g,S\left( \frac{%
u-v}{\lambda ^{2}}\right) g\rangle \right\} Y_{v}^{11}\left( \lambda \right)
a_{\lambda }\left( v\right) .
\end{eqnarray*}
If\ $v<u$, then $a_{\lambda }\left( u\right) $ can be commuted with $%
Y_{v}^{11}\left( \lambda \right) $ up to an $O_{u,v}\left( \lambda \right) $
error. Similarly, if $u<v$, then $a_{\lambda }^{\dag }\left( v\right) $ can
be commuted with $X_{u}^{11}\left( \lambda \right) $ up to an $O_{u,v}\left(
\lambda \right) $ error, thus the above equals 
\begin{eqnarray*}
&&\int_{0}^{t}du\,a_{\lambda }^{\dag }\left( u\right) X_{u}^{11}\left(
\lambda \right) \left\{ \int_{0}^{u}dva_{\lambda }^{\dag }\left( v\right)
Y_{v}^{11}\left( \lambda \right) a_{\lambda }\left( u\right) \right\}
a_{\lambda }\left( v\right) \\
&&+\int_{0}^{t}du\,a_{\lambda }^{\dag }\left( u\right) X_{u}^{11}\left(
\lambda \right) \left\{ \int_{0}^{u}dvO_{v,u}\left( \lambda \right)
a_{\lambda }\left( u\right) \right\} a_{\lambda }\left( v\right) \\
&&+\int_{0}^{t}dv\,a_{\lambda }^{\dag }\left( v\right) \left\{
\int_{0}^{v}dua_{\lambda }^{\dag }\left( u\right) X_{u}^{11}\left( \lambda
\right) a_{\lambda }\left( u\right) \right\} Y_{v}^{11}\left( \lambda
\right) a_{\lambda }\left( v\right) \\
&&+\int_{0}^{t}dv\int_{0}^{v}du\,a_{\lambda }^{\dag }\left( v\right)
a_{\lambda }\left( u\right) Y_{v}^{11}\left( \lambda \right) a_{\lambda
}\left( u\right) O_{u,v}\left( \lambda \right) \\
&&+\int_{0}^{t}du\int_{0}^{t}dv\,\frac{1}{\lambda ^{2}}\langle g,S\left( 
\frac{u-v}{\lambda ^{2}}\right) g\rangle a_{\lambda }^{\dag }\left( u\right)
X_{u}^{11}\left( \lambda \right) Y_{v}^{11}\left( \lambda \right) a_{\lambda
}\left( v\right) .
\end{eqnarray*}
In the limit $\lambda \rightarrow 0$ leads to 
\begin{eqnarray*}
&&\int_{0}^{t}X_{u}^{11}\left[ \int_{0}^{u}Y_{v}^{11}\otimes dA_{v}^{11}%
\right] \otimes dA_{u}^{11} \\
&&+\int_{0}^{t}\left[ X_{u}^{11}dA_{u}^{11}\right] \int_{0}^{u}Y_{v}^{11}%
\otimes dA_{v}^{11}+\int_{0}^{t}X_{u}^{11}Y_{u}^{11}\otimes dA_{u}^{11}.
\end{eqnarray*}
The last term is an Ito correction. Such a term results whenever the
pre-limit term $a_{\lambda }^{\dag }\left( v\right) a_{\lambda }\left(
u\right) $ is present and put into normal order. There are 4 such terms and
they lead to the usual quantum Ito correction 
\begin{equation*}
\lim_{\lambda \rightarrow 0^{+}}\left( X_{t}\left( \lambda \right)
-x_{0}\right) \left( Y_{t}\left( \lambda \right) -y_{0}\right) =\left(
X_{t}-x_{0}\right) \left( Y_{t}-y_{0}\right)
\end{equation*}
Thus the Ito calculus is picked up in the limit and the convergence for all
moments can be derived by induction.
\end{proof}

Here we have not attempted a most general statement which might be
formulated by defined processes as equivalence classes of sesquilinear forms
on the appropriate space of exponential vectors. Instead, we assumed
existence and uniqueness of solutions to begin with, however, situations
where this can be established will be presented in the next section.

\noindent \textbf{THEOREM 5.5.} Let $X_{t}\left( \lambda \right)
=x_{0}+\int_{0}^{t}du\left\{ a_{\lambda }^{\dag }\left( u\right)
X_{u}^{11}\left( \lambda \right) a_{\lambda }\left( u\right) +\cdots
\right\} $ with adaptable coefficients $X_{t}^{\alpha \beta }\left( \lambda
\right) $ such that $X_{t}\left( \lambda \right) ,X_{t}^{\alpha \beta
}\left( \lambda \right) $ converge respectively in weak matrix elements to $%
X_{t},X_{t}^{\alpha \beta }\left( \lambda \right) $ (uniformly for mixed
matrix elements) us above, then the following limits hold us quantum
Stratonovich integrals: 
\begin{eqnarray}
\int_{0}^{t}ds\,a_{\lambda }^{\dag }\left( s\right) a_{\lambda }\left(
s\right) X_{s}\left( \lambda \right) &\rightarrow &\int_{0}^{t}dA_{s}^{11}%
\overline{X_{s}},  \notag \\
\int_{0}^{t}ds\,X_{s}\left( \lambda \right) a_{\lambda }^{\dag }\left(
s\right) a_{\lambda }\left( s\right) &\rightarrow &\int_{0}^{t}\overline{%
X_{s}}dA_{s}^{11},  \notag \\
\int_{0}^{t}ds\,a_{\lambda }\left( s\right) X_{s}\left( \lambda \right)
&\rightarrow &\int_{0}^{t}dA_{s}^{01}\overline{X_{s}},  \notag \\
\int_{0}^{t}ds\,X_{s}\left( \lambda \right) a_{\lambda }^{\dag }\left(
s\right) &\rightarrow &\int_{0}^{t}\overline{X_{s}}dA_{s}^{10},  \notag \\
\int_{0}^{t}ds\,X_{s}\left( \lambda \right) a_{\lambda }\left( s\right)
&\rightarrow &\int_{0}^{t}\overline{X_{s}}dA_{s}^{01},  \notag \\
\int_{0}^{t}ds\,a_{\lambda }^{\dag }\left( s\right) X_{s}\left( \lambda
\right) &\rightarrow &\int_{0}^{t}dA_{s}^{10}\overline{X_{s}}.\, 
\TCItag{5.13}
\end{eqnarray}

\begin{proof}
The treatments of these limits are very similar, each involves at most one
reordering, and it is enough to work through just one of them: 
\begin{eqnarray*}
\int_{0}^{t}ds\,a_{\lambda }\left( s\right) X_{s}\left( \lambda \right)
&=&\int_{0}^{t}ds\,X_{s}\left( \lambda \right) a_{\lambda }\left( s\right) \\
&&+\int_{0}^{t}ds\int_{0}^{s}du\,[a_{\lambda }\left( s\right) ,a_{\lambda
}^{\dag }\left( u\right) X_{u}^{11}\left( \lambda \right) a_{\lambda }\left(
u\right) +\cdots ] \\
&=&\int_{0}^{t}ds\,X_{s}\left( \lambda \right) a_{\lambda }\left( s\right)
+\int_{0}^{t}ds\int_{0}^{s}du\,\,\frac{1}{\lambda ^{2}}\langle g,S\left( 
\frac{u-v}{\lambda ^{2}}\right) g\rangle \left\{ X_{u}^{11}\left( \lambda
\right) a_{\lambda }\left( u\right) +X_{u}^{10}\left( \lambda \right)
\right\} \\
&&+\int_{0}^{t}ds\int_{0}^{s}du\,\,a_{\lambda }^{\dag }\left( u\right)
O_{\lambda }a_{\lambda }\left( u\right) .
\end{eqnarray*}
Now, for $R_{\lambda }(u)$ the matrix elements of an adaptable process
between arbitrary collective exponential states, one has 
\begin{eqnarray*}
\lim_{\lambda \rightarrow 0^{+}}\int_{0}^{t}ds\int_{0}^{s}du\,\,\frac{1}{%
\lambda ^{2}}\langle g,S\left( \frac{u-v}{\lambda ^{2}}\right) g\rangle
R_{\lambda }\left( u\right) &=&\lim_{\lambda \rightarrow
0^{+}}\int_{0}^{t}ds\int_{0}^{s/\lambda ^{2}}d\tau \,\,\langle g,S\left(
\tau \right) g\rangle R_{\lambda }\left( s-\tau /\lambda ^{2}\right) \\
&=&\frac{1}{2}\int_{0}^{t}R\left( s\right) ds,
\end{eqnarray*}
where $R(u)$ is the associated limit; from the standard notion of weak
convergence in matrix elements from $\mathcal{H}_{0}\underline{\otimes }%
\mathcal{H}_{\lambda }$ to $\mathcal{H}_{0}\underline{\otimes }\mathcal{H}$
one sees that 
\begin{eqnarray*}
\lim_{\lambda \rightarrow 0^{+}}\int_{0}^{t}ds\,a_{\lambda }\left( s\right)
X_{s}\left( \lambda \right) &=&\int_{0}^{t}\,X_{s}\otimes dA_{s}^{01}+\frac{1%
}{2}\int_{0}^{t}\left( X_{s}^{11}\otimes dA_{s}^{10}+X_{s}^{10}\otimes
dA_{s}^{00}\right) \\
&=&\int_{0}^{t}dA_{s}^{01}\overline{X_{s}}.
\end{eqnarray*}
\end{proof}

\section{Asymptotic quantum stochastic limits. Uniformly convergent
situations}

The simplest way to bypass the adaptability ansatz introduced in the
previous section is to consider only linear ODES. It is possible, however,
to construct genuinely nonlinear examples starting from the linear case.
This, in fact, is very natural: all important dynamical equations in science
are nonlinear with the sole exception of the Schr\"{o}dinger equation, this
however leads to the Heisenberg evolution which is generally nonlinear. The
program of this section is as follows. It is first of all shown that
solutions $U_{t}\left( \lambda \right) $ to linear Wick-ordered ODES
converge without adopting an adaptability ansatz. If the limit process $%
U_{t} $, is unitary, then it is shown that processes $X_{t}=U_{t}^{\dag
}x_{0}U_{t} $, for some $x_{0}\in \mathcal{B}\left( \mathcal{H}_{0}\right) $%
, satisfy nonlinear QSDEs. In a sense, this class of solutions is the most
interesting as they provide examples of quantum dynamical variables having a
symplectic evolution.

\bigskip

\noindent \textbf{THEOREM 6.1.} Let $x_{0},C_{\alpha \beta }\in \mathcal{B}%
\left( \mathcal{H}_{0}\right) $ for $\alpha ,\beta \in \left\{ 0,1\right\} $
and let $X_{t}\left( \lambda \right) $ be the solution to the linear ODE

\begin{equation}
\dot{X}_{t}\left( \lambda \right) =a_{\lambda }^{\dag }\left( t\right)
C_{11}X_{t}\left( \lambda \right) a_{\lambda }\left( t\right) +a_{\lambda
}^{\dag }\left( t\right) C_{10}X_{t}\left( \lambda \right)
+C_{01}X_{t}\left( \lambda \right) a_{\lambda }\left( t\right)
+C_{00}X_{t}\left( \lambda \right)  \tag{6.1}
\end{equation}
with $X_{0}\left( \lambda \right) =x_{0}$, then for all $u,v\in \mathcal{H}%
_{0}$ and $f,h\in \mathcal{S}$%
\begin{equation*}
\lim_{\lambda \rightarrow 0^{+}}\langle u\underline{\otimes }\Psi \left(
f,\lambda \right) ,X_{t}\left( \lambda \right) \,v_{2}\underline{\otimes }%
\Psi \left( k,\lambda \right) \rangle _{\mathcal{H}_{0}\underline{\otimes }%
\mathcal{H}}=\langle u\underline{\otimes }\Psi \left( f\right) ,X_{t}\,v%
\underline{\otimes }\Psi \left( k\right) \rangle _{\mathcal{H}_{0}\underline{%
\otimes }\mathcal{H}},
\end{equation*}
where $X_{t}$ is the solution to the QSDE 
\begin{equation}
dX_{t}=\sum_{\alpha ,\beta }C_{\alpha \beta }X_{t}\otimes dA_{t}^{\alpha
\beta };\quad X_{0}=x_{0}.  \tag{6.2}
\end{equation}

\begin{proof}
Let $a_{\lambda }^{\alpha }\left( t\right) =a_{\lambda }\left( t\right) $
for $\alpha =1$, and $=1$ for $\alpha =0$. Eq. (6.1) can be re-written as 
\begin{equation*}
\dot{X}_{t}\left( \lambda \right) =\sum_{\alpha ,\beta }a_{\lambda }^{\alpha
}\left( t\right) ^{\dag }C_{\alpha \beta }X_{t}\left( \lambda \right)
a_{\lambda }^{\beta }\left( t\right) .
\end{equation*}
As this is linear, the formal iterative series expansion exists, 
\begin{equation*}
X_{t}\left( \lambda \right) =x_{0}+\sum_{n=1}^{\infty }X_{t}^{\left(
n\right) }\left( \lambda \right)
\end{equation*}
where 
\begin{eqnarray*}
X_{t}^{\left( n\right) }\left( \lambda \right) &=&\sum_{\alpha _{1},\beta
_{1},\cdots ,\alpha _{n},\beta _{n}}\int_{t>t_{n}>\cdots >t_{1}\geq
0}dt_{1}\cdots dt_{n} \\
&&\times \left( a_{\lambda }^{\alpha _{1}}\left( t_{1}\right) \cdots
a_{\lambda }^{\alpha _{n}}\left( t_{n}\right) \right) ^{\dag }C_{\alpha
_{1}\beta _{1}}\cdots C_{\alpha _{n}\beta _{n}}x_{0}\,a_{\lambda }^{\beta
_{1}}\left( t_{1}\right) \cdots a_{\lambda }^{\beta _{n}}\left( t_{n}\right)
\end{eqnarray*}
In particular, 
\begin{eqnarray}
\langle u\underline{\otimes }\Psi \left( f,\lambda \right) ,X_{t}\left(
\lambda \right) \,v_{2}\underline{\otimes }\Psi \left( k,\lambda \right)
\rangle _{\mathcal{H}_{0}\underline{\otimes }\mathcal{H}}= &&\sum_{\alpha
_{1},\beta _{1},\cdots ,\alpha _{n},\beta _{n}}\int_{t>t_{n}>\cdots
>t_{1}\geq 0}dt_{1}\cdots dt_{n}\langle u,C_{\alpha _{1}\beta _{1}}\cdots
C_{\alpha _{n}\beta _{n}}x_{0}v\rangle _{\mathcal{H}_{0}}  \notag \\
&&\times f_{\lambda }^{\alpha _{1}}\left( t_{1}\right) ^{\ast }\cdots
f_{\lambda }^{\alpha _{n}}\left( t_{n}\right) ^{\ast }k_{\lambda }^{\beta
_{1}}\left( t_{1}\right) \cdots k_{\lambda }^{\beta _{n}}\left( t_{n}\right)
\TCItag{6.3}
\end{eqnarray}
where $f_{\lambda }^{\alpha }\left( t\right) :=\int_{0}^{\infty }f\left(
t+\lambda ^{2}u\right) \langle S\left( u\right) g,g\rangle du$ for $\alpha
=1 $, and $=1$ for $\alpha =0$, etc.

Expression (6.3) is bounded by $c_{0}\left\| u\right\| \left\| v\right\|
\left( c\langle \Gamma g,g\rangle \left\| f\right\| _{\infty }\left\|
k\right\| _{\infty }t\right) ^{n}$\ where $c:=\max \left\| C_{\alpha \beta
}\right\| $. The series expansion is therefore uniformly convergent, with
(6.3) converging to 
\begin{eqnarray*}
&&\sum_{\alpha _{1},\beta _{1},\cdots ,\alpha _{n},\beta
_{n}}\int_{t>t_{n}>\cdots >t_{1}\geq 0}dt_{1}\cdots dt_{n}\langle
u,C_{\alpha _{1}\beta _{1}}\cdots C_{\alpha _{n}\beta _{n}}x_{0}v\rangle _{%
\mathcal{H}_{0}} \\
&&\times f^{\alpha _{1}}\left( t_{1}\right) ^{\ast }\cdots f^{\alpha
_{n}}\left( t_{n}\right) ^{\ast }k^{\beta _{1}}\left( t_{1}\right) \cdots
k^{\beta _{n}}\left( t_{n}\right) ,
\end{eqnarray*}
where $f^{1}\left( t\right) :=\langle S_{+}g,g\rangle f\left( t\right) $%
%TCIMACRO{\UNICODE{0xa8}}%
%BeginExpansion
\"{}%
%EndExpansion
and $f^{0}\left( t\right) :=1$. Resumming gives the correct matrix element
for the process $X_{t}$ described in the statement of the theorem.
\end{proof}

The expediency of Theorem 6.1 compared to those employed in the papers of
Accardi, Frigerio and Lu [9] comes about from the fact that here the limit
is anticipated via white noise operators. The mechanism is transparent
because the pre-limit and limit representations are Wick ordered. The
difficulty encountered there was that the prelimit fields were Weyl ordered
(which is generally the ordering natural to equations of elementary physics)
and so enormous efforts were spent in re-ordering and the subsequent
identification and treatment of negligible terms. Once it is known that the
process converges, however, the adaptability ansatz can be dispensed with
for the coefficients of the ODE. Part (ii) of Theorem 5.4 affirms that the
multi-moment convergence of the process occurs. Therefore the following
conclusion is reached.

\bigskip

\noindent \textbf{THEOREM 6.2.} Let $X_{t}\left( \lambda \right) $ be the
solution of the linear ODE (6.1) then $X_{t}\left( \lambda \right) $
converges weakly to the solution of the QSDE (6.2) in matrix elements.

\bigskip

The problem of obtaining a unitary process $U_{t}$\ obeying a linear QSDE
has been tackled [12, 171 and can be summarized in the next theorem. Here $%
a^{\sharp }\left( t\right) =a^{\sharp }\left( t,g\right) $ with $\langle
S_{+}g,g\rangle $ taken as $\kappa =\frac{1}{2}\gamma +i\sigma $. The CCR
for the noise is then $[a(t),a^{\dag }(s)]=\kappa \delta _{+}\left(
t-s\right) +\kappa _{\ast }\delta _{-}\left( t-s\right) $. With the
identifications $dA_{t}^{11}=a^{\dag }\left( t\right) a\left( t\right) dt$, $%
dA_{t}^{10}=a^{\dag }\left( t\right) dt$, $dA_{t}^{01}=a\left( t\right) dt$
and $dA_{t}^{00}=dt$ one is led to the nontrivial component of the lto table
being $dA_{1}^{\alpha 1}dA_{t}^{1\beta }=\gamma dA_{t}^{\alpha \beta }$.

\bigskip

\noindent \textbf{THEOREM 6.3.} The general unitary process [23] $\left(
U_{t}\right) _{t\geq 0}$ driven by white noise processes $a^{\sharp }\left(
t\right) $ is given by ($U_{0}=1$) 
\begin{eqnarray}
\dot{U}_{t} &=&\left[ 1,\frac{1}{\sqrt{\gamma }}a^{\dag }\left( t\right) %
\right] \left[ 
\begin{array}{cc}
-iH-\frac{1}{2}L^{\dag }L & -L^{\dag }W \\ 
L & W-1
\end{array}
\right] U_{t}\left[ 
\begin{array}{c}
1 \\ 
\frac{1}{\sqrt{\gamma }}a\left( t\right)
\end{array}
\right]  \TCItag{6.4} \\
&=&-i\left[ Ea^{\dag }\left( t\right) a\left( t\right) +Fa^{\dag }\left(
t\right) +F^{\dag }a\left( t\right) +G\right]  \TCItag{6.5}
\end{eqnarray}
where $E,H,G$ are self-adjoint and $W$ is unitary on $\mathcal{H}_{0}$, and
the coeficients of the Ito QSDE (6.4) are related to those of the
Stratonovich QSDE (6.5) by the relations 
\begin{equation}
W=\frac{1-i\kappa ^{\ast }E}{1+i\kappa E},\quad L=-i\sqrt{\gamma }\left(
1+i\kappa E\right) ^{-1}F,\quad H=G+F^{\dag }\frac{\sigma -\left| \kappa
\right| ^{2}E}{1+\left| \kappa \right| ^{2}E}F.  \tag{6.6}
\end{equation}

One notes that the Stratonovich QSDE takes the form $\dot{U}_{t}=-i\Upsilon
_{t}U_{t}$, where $\Upsilon _{t}=Ea^{\dag }\left( t\right) a\left( t\right)
+Fa^{\dag }\left( t\right) +F^{\dag }a\left( t\right) +G$ can be interpreted
as a quantum stochastic Hamiltonian.

\textbf{\noindent THEOREM 6.4.} Let $E,F,G\in \mathcal{B}\left( \mathcal{H}%
_{0}\right) $ with $E$ and $G$ self-adjoint then the family of processes $%
U_{\lambda }\left( t\right) $ satisfying the ODES 
\begin{equation*}
\dot{U}_{t}\left( \lambda \right) =-i\Upsilon _{t}\left( \lambda \right)
U_{t}\left( \lambda \right) ,\quad U_{0}\left( \lambda \right) =1,
\end{equation*}
with 
\begin{equation*}
\Upsilon _{t}\left( \lambda \right) =Ea_{t}^{\dag }\left( \lambda \right)
a_{t}\left( \lambda \right) +Fa_{t}^{\dag }\left( \lambda \right) +F^{\dag
}a_{t}\left( \lambda \right) +G,
\end{equation*}
converges weakly in matrix elements to the unitary quantum stochastic
process described in Theorem 6.3.

\begin{proof}
One begins by noting that 
\begin{eqnarray}
\left[ a_{\lambda }\left( t\right) ,U_{t}\left( \lambda \right) \right] &=&-i%
\left[ a_{\lambda }\left( t\right) ,\int_{0}^{t}ds\,\Upsilon _{s}\left(
\lambda \right) U_{s}\left( \lambda \right) \right]  \notag \\
&=&-i\int_{0}^{t}ds\frac{1}{\lambda ^{2}}\langle g,S\left( \frac{t-s}{%
\lambda ^{2}}\right) g\rangle \left\{ Ea_{s}\left( \lambda \right)
+F\right\} U_{s}\left( \lambda \right)  \notag \\
&\equiv &-i\langle S_{+}g,g\rangle \left\{ Ea_{t}\left( \lambda \right)
+F\right\} U_{t}\left( \lambda \right) +O\left( \lambda \right) . 
\TCItag{6.7}
\end{eqnarray}
Rearranging the ODE from Weyl to Wick ordered form gives 
\begin{eqnarray*}
\dot{U}_{t}\left( \lambda \right) &=&\frac{1}{\gamma }a_{\lambda }^{\dag
}\left( t\right) \left( W-1\right) U_{t}a_{\lambda }\left( t\right) +\frac{1%
}{\sqrt{\gamma }}a_{\lambda }^{\dag }\left( t\right) LU_{t} \\
&&-\frac{1}{\sqrt{\gamma }}L^{\dag }WU_{t}a_{\lambda }\left( t\right)
-\left( \frac{1}{2}L^{\dag }L+iH\right) U_{t}+O\left( \lambda \right) ;
\end{eqnarray*}
one recognizes the pre-limit form of (6.4) and the results of the previous
section imply convergence.
\end{proof}

\bigskip

The weak convergence in matrix elements, however, gives much more.

\noindent \textbf{THEOREM 6.5.} Let $x_{o}\in \mathcal{B}\left( \mathcal{H}%
_{0}\right) $, and set $X_{t}\left( \lambda \right) =U_{t}^{\dag }\left(
\lambda \right) x_{0}U_{t}\left( \lambda \right) $, then $X_{t}\left(
\lambda \right) $ converges weakly in matrix elements to the quantum
stochastic process $X_{t}=U_{t}^{\dag }x_{0}U_{t}$.

\bigskip

Given $X_{t}=U_{t}^{\dag }x_{0}U_{t}$, then one has the stochastic evolution
equations 
\begin{equation}
\dot{X}_{t}=\dot{U}_{t}^{\dag }x_{0}U_{t}+U_{t}^{\dag }x_{0}\dot{U}_{t}=%
\frac{1}{i}\left[ X_{t},\Theta _{t}\right]  \tag{6.8}
\end{equation}
where $\Theta _{t}:=U_{t}^{\dag }\Upsilon U_{t}$. The second part takes the
form of a stochastic Heisenberg equation. It is a straightforward
calculation to show, either by converting to the Hudson-Parthasarathy
calculus using the quantum Ito formula and inverting back or by the now
standard manipulations using the white noise calculus to put to Wick order,
that the QSDE for $X_{t}$, is 
\begin{eqnarray}
\dot{X}_{t} &=&\frac{1}{\gamma }a^{\dag }\left( t\right) \left( W_{t}^{\dag
}X_{t}W_{t}-X_{t}\right) a\left( t\right)  \notag \\
&&+\frac{1}{\sqrt{\gamma }}a^{\dag }\left( t\right) W_{t}^{\dag
}[X_{t},L_{t}]-\frac{1}{\sqrt{\gamma }}\left[ X_{t},L_{t}^{\dag }\right]
W_{t}U_{t}a_{\lambda }\left( t\right)  \notag \\
&&-\frac{1}{2}\left[ L_{t}^{\dag },X_{t}\right] L_{t}-\frac{1}{2}L_{t}^{\dag
}\left[ X_{t},L_{t}\right] -i\left[ X_{t},H_{t}\right] .  \TCItag{6.9}
\end{eqnarray}
The Stratonovich version of this will now be computed.

\noindent \textbf{LEMMA 6.6.} Let $U_{t}$ be the solution to the QSDEs (6.4)
and (6.5), then one can perform the following commutations under the
integral: 
\begin{eqnarray*}
U_{t}^{\dag }a^{\dag }\left( t\right) &=&\left\{ a^{\dag }\left( t\right)
U_{t}^{\dag }+i\kappa ^{\ast }U_{t}^{\dag }F^{\dag }\right\} \left(
1-i\kappa ^{\ast }E\right) ^{-1}, \\
U_{t}^{\dag }a\left( t\right) &=&\left\{ a\left( t\right) U_{t}^{\dag
}-i\kappa U_{t}^{\dag }F\right\} \left( 1+i\kappa E\right) ^{-1}, \\
U_{t}^{\dag }a^{\dag }\left( t\right) a\left( t\right) &=&\{a^{\dag }\left(
t\right) a\left( t\right) U_{t}^{\dag }-i\kappa a^{\dag }\left( t\right)
U_{t}^{\dag }\left( 1-i\kappa ^{\ast }E\right) ^{-1}F \\
&&+i\kappa ^{\ast }a\left( t\right) U_{t}^{\dag }\left( 1+i\kappa E\right)
^{-1}F^{\dag }+|\kappa |^{2}U_{t}^{\dag }F\left( 1-i\kappa ^{\ast }E\right)
^{-1}F^{\dag } \\
&&+|\kappa |^{2}U_{t}^{\dag }F^{\dag }\left( 1-i\kappa ^{\ast }E\right)
^{-1}F\}\left( 1-2\sigma E\right) ^{-1}.
\end{eqnarray*}

\begin{proof}
Before demonstrating the proof it is important to point out that (6.12)
cannot be obtained from (6.10) and (6.11); this is due to the
afore-mentioned nonassociativity (that is, the product of Stratonovich
integrands is not defined as the Stratonovich integrand of the products). To
be consistent, the left-hand side of (6.12) should be written as
U/at(t)a(t); the symbol U/at(t)Z(t) has not been defined. However, the
underbar and overbar notation has been dropped for convenience. To prove
(6.10) one has directly 
\begin{eqnarray*}
\left[ a\left( t\right) ,U_{t}\right] &=&-i\int_{0}^{t}\left[ a\left(
t\right) ,\left( Ea^{\dag }\left( t\right) a\left( t\right) +Fa^{\dag
}\left( t\right) +F^{\dag }a\left( t\right) +G\right) U_{s}\right] ds \\
&=&-i\int_{0}^{t}\left( Ea\left( s\right) +F\right) \left\{ \kappa \delta
_{+}\left( t-s\right) +\kappa _{\ast }\delta _{-}\left( t-s\right) \right\}
U_{s} \\
&=&-i\kappa \left( Ea\left( s\right) +F\right) U_{t}.
\end{eqnarray*}
The right-hand side is not properly normal ordered; however, it can be
rearranged to give 
\begin{equation}
a\left( t\right) U_{t}=\left( 1+i\kappa E\right) ^{-1}\left\{ U_{t}a\left(
t\right) -i\kappa FU_{t}\right\} .  \tag{6.13}
\end{equation}
Note that this can be deduced from the pre-limit expressions and corresponds
then to (6.7). Eq. (6.10) is the conjugate equation to (6.13). The
derivation of (6.11) is similar.

To derive (6.12), note first of all that 
\begin{eqnarray}
\left[ a^{\dag }\left( t\right) a\left( t\right) ,\Upsilon _{t}\right]
&=&Ea^{\dag }\left( t\right) a\left( s\right) \left\{ \kappa \delta
_{+}\left( t-s\right) +\kappa _{\ast }\delta _{-}\left( t-s\right) \right\}
-Ea^{\dag }\left( s\right) a\left( t\right) \left\{ \kappa \delta _{+}\left(
s-t\right) +\kappa _{\ast }\delta _{-}\left( s-t\right) \right\}  \notag \\
&&+Fa^{\dag }\left( t\right) \left\{ \kappa \delta _{+}\left( t-s\right)
+\kappa _{\ast }\delta _{-}\left( t-s\right) \right\} -F^{\dag }a\left(
t\right) \left\{ \kappa \delta _{+}\left( s-t\right) +\kappa _{\ast }\delta
_{-}\left( s-t\right) \right\} .  \TCItag{6.14}
\end{eqnarray}
Therefore, 
\begin{eqnarray}
\left[ a^{\dag }\left( t\right) a\left( t\right) ,U_{t}\right]
&=&-i-i\int_{0}^{t}\left[ a^{\dag }\left( t\right) a\left( t\right)
,\Upsilon _{s}U_{s}\right] ds  \notag \\
&=&-iEa^{\dag }\left( t\right) a\left( t\right) \left( \kappa -\kappa ^{\ast
}\right) U_{t}-i\kappa Fa^{\dag }\left( t\right) U_{t}+i\kappa ^{\ast
}F^{\dag }a\left( t\right) U_{t}.  \TCItag{6.15}
\end{eqnarray}

Again (6.15) is not wholly ordered, however, whereas the manipulations are
not associative, they are linear and so $a^{\sharp }\left( t\right) U_{t}$
can be replaced using (6.13), etc. This will lead to the conjugate equation
to (6.12).
\end{proof}

The first form of the Heisenberg equation (6.8) can be written as 
\begin{equation}
\dot{X}_{t}=U_{t}^{\dag }\left\{ a^{\dag }\left( t\right) a\left( t\right) 
\frac{1}{i}\left[ x_{0},E\right] +a^{\dag }\left( t\right) \frac{1}{i}\left[
x_{0},F\right] +\frac{1}{i}\left[ x_{0},F^{\dag }\right] a\left( t\right) +%
\frac{1}{i}\left[ x_{0},G\right] \right\} U_{t}.  \tag{6.16}
\end{equation}
The conversion to Stratonovich form (left-handed) can now be made using the
lemma. One obtains 
\begin{eqnarray}
\dot{X}_{t} &=&a^{\dag }\left( t\right) a\left( t\right) \left( 1-2\sigma
E_{t}\right) ^{-1}\frac{1}{i}\left[ X_{t},E_{t}\right]  \notag \\
&&+a^{\dag }\left( t\right) \left\{ -i\kappa \left( 1-i\kappa ^{\ast
}E_{t}\right) ^{-1}F_{t}\left( 1-\sigma E_{t}\right) ^{-1}\frac{1}{i}\left[
X_{t},E_{t}\right] +\left( 1-i\kappa ^{\ast }E_{t}\right) ^{-1}\frac{1}{i}%
\left[ X_{t},F_{t}\right] \right\}  \notag \\
&&+a\left( t\right) \left\{ i\kappa ^{\ast }\left( 1+i\kappa E_{t}\right)
^{-1}F_{t}^{\dag }\left( 1-\sigma E_{t}\right) ^{-1}\frac{1}{i}\left[
X_{t},E_{t}\right] +\left( 1+i\kappa E_{t}\right) ^{-1}\frac{1}{i}\left[
X_{t},F_{t}^{\dag }\right] \right\}  \notag \\
&&+\{\frac{1}{i}\left[ X_{t},H_{t}\right] +\left| \kappa \right|
^{2}F_{t}\left( 1+i\kappa E_{t}\right) ^{-1}F_{t}^{\dag }\left( 1-2\sigma
E_{t}\right) ^{-1}\frac{1}{i}\left[ X_{t},E_{t}\right]  \notag \\
&&+\left| \kappa \right| ^{2}F_{t}^{\dag }\left( 1-i\kappa ^{\ast
}E_{t}\right) ^{-1}F_{t}\left( 1-2\sigma E_{t}\right) ^{-1}\frac{1}{i}\left[
X_{t},E_{t}\right]  \notag \\
&&+i\kappa ^{\ast }F_{t}^{\dag }\left( 1-i\kappa ^{\ast }E_{t}\right) ^{-1}%
\frac{1}{i}\left[ X_{t},F_{t}\right] -\kappa F_{t}\left( 1+i\kappa
E_{t}\right) ^{-1}\frac{1}{i}\left[ X_{t},F_{t}^{\dag }\right] . 
\TCItag{6.17}
\end{eqnarray}

Eqs. (6.9) and (6.17) are equivalent and give the most general Heisenberg
equation for this situation.

\bigskip

\noindent \textbf{Remarks}

The dynamical evolutions considered in this section are broad enough to
include the classical SDEs considered in Sections 2 and 3.

Let $x_{0},p_{0}$ be canonically conjugate variables on $\mathcal{H}_{0}$.
Here the specification $\kappa =\frac{1}{2}$, $W=1$, $L=-i\tilde{\sigma}%
\left( x_{0}\right) p_{0}$, and $G=\frac{1}{2}\left( \tilde{v}\left(
x_{0}\right) p_{0}+p\tilde{v}\left( x_{0}\right) \right) $ leads to (6.9)
being the diffusion described in (4.8b). The QSDE has unbounded operator
coefficients, but this can be dealt with using standard techniques [24].
Here $E=0,F=p_{0}\tilde{\sigma}\left( x_{0}\right) $, and $H=G$, and so
(6.17) reduces to 
\begin{equation*}
\end{equation*}
Here $P_{t}=U_{t}^{\dag }p_{0}U_{t}$ and since the evolution is unitary, $%
[X_{t},P_{t}]=i$ for each $t\geq 0$. The last term in (6.18) is therefore $%
\frac{1}{2}i\left[ \tilde{\sigma}\left( X_{t}\right) ,P_{t}\right] \tilde{%
\sigma}\left( X_{t}\right) =-\frac{i}{2}\tilde{\sigma}^{\prime }\left(
X_{t}\right) \tilde{\sigma}\left( X_{t}\right) $ which is the correct drift
term, and so (6.17) reduces to (4.8a). The classical relations are therefore
recovered.

Likewise, to obtain Eqs. (4.9) for the Poisson driven processes, one sets $%
\kappa =\frac{1}{2}$ and takes $E=F=F^{\dag }=H-\frac{1}{2}%
(p_{0}v(x_{0})+v(x_{0})p_{0})=\frac{1}{2}(p_{0}\mu (x_{0})+\mu (x_{0})p_{0})$%
. \ In this case $W$ can be represented as the change-of-variable operator ($%
Wf)(x)==\sqrt{\frac{du}{dx}}f\left( u\left( x\right) \right) $ where $%
u=u(x):=x+\tilde{\mu}(x)$. In this case $W^{\dag }x_{0}W=x_{0}+\tilde{\mu}%
\left( x_{0}\right) $, with $x_{0}$ interpreted as the multiplication by $x$
operator on $\mathcal{H}_{0}:=L^{2}\left( \mathbb{R}\right) $.

Acknowledgment

This work was supported by the Irish Higher Education Authority EU
Presidency Research Fellowship Program and the author would like to thank
Professor D. Heffernan (Maynooth) for many stimulating discussions while
writing this paper. The author also acknowledges many valuable comments made
by the referee which have lead to an improvement to the original version.

REFERENCES

[1] A. D. Viencel and M. I. Freidlin: Fluctuations In Dynamical Systems
Subjected to Small Random Perturbations, Nauka, Moscow 1979 (in Russian)

[2] K. Sobczyk: Stochastic Differential Equations, Mathematics in
Applications, Vol. 40, Kluwer, Dordrecht 1991.

[3] R. Z. Khasminskii: A limit theorem for the solution of a differential
equation with random right hand side, Teoria Vieroyatn. Prim. 11, No. 2
(1966).

[4] E. Wong and M. Zakai: ht. J. Eng. Sci. 3 (1965), 213-229.

[5] R. L. Stratonovich: SIAM J. Control 4 (1966), 362-371.

[6] L. van Hove: Physica 21 (1955), 617-640.

[7] W. H. Louisell: Quantum Statistical Properties of Radiation, Wiley
Classics Library Edition, New York 1990.

[8] L. Accardi, J. Cough and Y. G. Lu: Rep. Math. Phys. 36 (2/3) (1995),
155-187.

[9] L. Accardi, A. Frigerio and Y. G. Lu: Commun. Math. Phys. 131 (1990)
537-570.

[10 J. Gough, C. R. Acad, Sci. Paris 326, Serie I (1998), 981-985.

[11] C. W. Gardiner and M. J. Collet: Phys. Rev. A 31 (1985), 3761-3774.

[12] R. L. Hudson and K. R. Parthasarathy: Commun. Math. Phys. 93 (1984),
301-323.

[ 131 R. L. Hudson and R. Streater: Phys. Left. A 277 (1981).

[14] A. M. Chebotarev: Mat. Zametki 60, No. 5 (1996), 726-750.

[15] P. Talkner: Ann. Phys. 167 (1980), 390.

[16] J. Gough: Causal structure of quantum stochastic integrators,
Theoretical Mathematical Physics 111, No. 2, May (1997), 218-233 (with
English translation 563-575).

[17] J. Gough: Non-commutative Ito and Stratonovich noise and stochastic
evolutions, Theoretical and Mathematical Physics 113, No. 2, November
(1997), 276-284 (with English translation 1431-1437).

[18] J. Gough: The Stratonovich interpretation of quantum stochastic
approximations, Journ. Potential Analysis, Volume 11, Issue 3, pp 213-233,
November 1999

[19] I. I. Gihman and A. V. Skorohod: Stochastic Differential Equations,
Springer, Berlin 1974.

[20] K. R. Parthasarathy: An Introduction to Quantum Stochastic Calculus,
Monographs in Mathematics, Birkhauser, Base1 1992.

[21] J. Dieudonne: Foundations of Modern Analysis, Academic Press, New York
1968.

[22] F. A. Berezin: Vsp. Fiz. Nauk. 132 (November 1980), 497-548.

[23] The Ito QSDEs with $W=-1$ are not convertible to standard ($\kappa =1$)
Stratonovich form as the first of Eq. (6.6) is noninvertible. The reflection
process is an example.

[24] K. B. Sinha: Quantum Stochastic Calculus and Applications-A Review,
Probability Towards 2000, Springer, New York 1998.

\end{document}